\documentclass[12pt]{article}
\usepackage{a4}
\usepackage{amsmath}
\usepackage{array}
\usepackage{amsfonts}
\newcommand{\ga}{\alpha}
\newcommand{\gb}{\beta}
\renewcommand{\gg}{\gamma}
\newcommand{\gd}{\delta}
\renewcommand{\ge}{\epsilon}

\newcommand{\gf}{\phi}

\newcommand{\gx}{\xi}
\newcommand{\gm}{\mu}
\newcommand{\gn}{\nu}

\newcommand{\gl}{\lambda}

\newcommand{\gth}{\theta}
\newcommand{\gs}{\sigma}

\newcommand{\go}{\omega}

\newcommand{\gp}{\pi}
\newcommand{\gps}{\psi}
\newcommand{\get}{\eta}
\newcommand{\gch}{\chi}
\newcommand{\Bgf}{{\boldsymbol \phi}}

\newcommand{\Bgps}{{\boldsymbol \psi}}

\newcommand{\gG}{\Gamma}
\newcommand{\gD}{\Delta}
\newcommand{\gF}{\Phi}

\newcommand{\gS}{\Sigma}
\newcommand{\gTh}{\Theta}

\newcommand{\cD}{{\cal D}}

\newcommand{\cF}{{\cal F}}

\newcommand{\cK}{{\cal K}}
\newcommand{\cL}{{\cal L}}
\newcommand{\cM}{{\cal M}}

\newcommand{\cO}{{\cal O}}

\newcommand{\cR}{{\cal R}}

\newcommand{\fJ}{{\frak J}}
\newcommand{\fK}{{\frak K}}

\newcommand{\tK}{{\tilde K}}

\newcommand{\tM}{{\tilde M}}

\newcommand{\siA}{{ A^\dashv}}
\newcommand{\siB}{{ B^\dashv}}
\newcommand{\siC}{{ C^\dashv}}
\newcommand{\siD}{{ D^\dashv}}

\newcommand{\siga}{{ \alpha^\dashv}}
\newcommand{\sigb}{{\beta^\dashv}}
\newcommand{\sigg}{{\gamma^\dashv}}
\newcommand{\sigd}{{\delta^\dashv}}

\newcommand{\uga}{{\underline \alpha}}
\newcommand{\ugb}{{\underline\beta}}

\newcommand{\tget}{{\tilde\eta}}
\newcommand{\tgch}{{\tilde\chi}}
\newcommand{\tgG}{{\tilde\Gamma}}

\newcommand{\bk}{{\bar k}}

\newcommand{\bp}{{\bar p}}

\newcommand{\bv}{{\bar v}}

\newcommand{\bx}{{\bar x}}

\newcommand{\bz}{{\bar z}}
\newcommand{\bA}{{\bar A}}

\newcommand{\bF}{{\bar F}}

\newcommand{\bL}{{\bar L}}

\newcommand{\bR}{{\bar R}}
\newcommand{\bS}{{\bar S}}
\newcommand{\bT}{{\bar T}}

\newcommand{\bX}{{\bar X}}

\newcommand{\bga}{{\bar \alpha}}
\newcommand{\bgb}{{\bar\beta}}
\newcommand{\bgg}{{\bar\gamma}}
\newcommand{\bgd}{{\bar\delta}}

\newcommand{\bgG}{{\bar\Gamma}}

\newcommand{\Tr}{\mbox{Tr}}
\newcommand{\tr}{\text{tr}}

\newcommand{\Id}{\text{\small 1}\hspace{-3.5pt}\text{1}}

\newcommand{\Slashed}{\hspace{-1.3ex}/\hspace{.2ex}}
\newcommand{\lra}{\longrightarrow}
\newcommand{\Lra}{\Longrightarrow}

\newcommand{\der}{\partial}

\newcommand{\Der}{D}
\newcommand{\sDer}{\Der\Slashed}

\newcommand{\inv}{^{-1}}

\newcommand{\qand}{\quad \text{and} \quad}
\newcommand{\pset}[2]{^{#1} \! {#2}}
\newcommand{\lw}[1]{^{\;}_{#1}}
\newcommand{\nit}{\noindent}
\newcommand{\nl}{\newline}
\newcommand{\np}{\newpage}
\newcommand{\dsp}{\displaystyle}
\newcommand{\scp}{\scriptstyle}
\newcommand{\sscp}{\scriptscriptstyle}
\newcommand{\ct}{\cite}
\newcommand{\bit}{\bibitem}
\newcommand{\lh}{\left(}
\newcommand{\rh}{\right)}

\newcommand{\undr}[1]{{\underline{#1}}}
\newcommand{\ovr}[1]{{\overline{#1}}}

\newcommand{\non}{\nonumber}
 
\newcommand{\labl}[1]{\label{#1}}
\newcommand{\half}{\frac 12 }

\newcommand{\Kh}{K\"{a}hler}

\newcommand{\Intr}{\mathbb{Z}}
\newcommand{\Cplx}{\mathbb{C}}
\newcommand{\Real}{\mathbb{R}}
\newcommand{\beq}{\begin{gather}}
\newcommand{\eeeeeq}{\end{gather}}
\newcommand{\barr}{\begin{array}}
\newcommand{\earr}{\end{array}}
\newcommand{\equ}[1]{\begin{gather} #1 
\end{gather}}
\newcommand{\equa}[1]{\begin{align} #1 
\end{align}}

\newcommand{\arry}[3]{\renewcommand{\arraystretch}{#2}
\begin{array}{#1} #3 \end{array}}

\newcommand{\mtrx}[1]{\begin{matrix} #1 \end{matrix}}
\newcommand{\pmtrx}[1]{\begin{pmatrix} #1 \end{pmatrix}}
\newcommand{\cntr}[1]{\begin{center} #1 \end{center}}
\newcommand{\capt}[1]{\caption{{\small #1}}}
\newcounter{oldcounter}

\newcommand{\be}{\begin{equation}}
\newcommand{\ee}{\end{equation}}
\newcommand{\bea}{\begin{eqnarray}}
\newcommand{\eea}{\end{eqnarray}}

\newcommand{\commentary}[1]{}

\begin{document}
%
%
\pagestyle{empty} 
 
\begin{flushright} 
NIKHEF/00-026
\end{flushright}  
 
\begin{center} 
{\Large {\bf Construction and analysis of anomaly-free 
supersymmetric $\boldsymbol{SO(2N)/U(N)}$ 
$\boldsymbol{\sigma}$-models}}\\ 
\vspace {5ex} 
 
{\large S.\ Groot Nibbelink, T.S.\ Nyawelo}\\
\vspace{2ex} 
{\large and J.W.\ van Holten}\\ 
\vspace{3ex} 
 
{\large NIKHEF, P.O.\ Box 41882,}\\ 
\vspace{3ex} 
 
{\large 1009 DB, Amsterdam NL} 
\vspace{5ex} 

\today

\end{center} 
\vspace{15ex} 
 
\nit 
{\small 
{\bf Abstract} This paper discusses a procedure for the 
consistent coupling of gauge- and matter superfields to 
supersymmetric sigma-models on symmetric coset spaces of 
K\"{a}hler type. We exhibit the finite isometry transformations 
and the corresponding K\"{a}hler transformations. These lead to 
the construction of a generalized type of Killing potentials. 
In certain cases a charge quantization condition needs to be 
imposed to guarantee the global existence of a line bundle 
on a coset space. The results are applied to the explicit 
construction of sigma-models on cosets $SO(2N)/U(N)$. Only a 
finite number of these models can consistently incorporate 
matter in representations descending from the spinorial 
representations of $SO(2N)$. We investigate in detail
some aspects of the vacuum structure of the gauged 
$SO(10)/U(5)$ theory, with surprising results: the fully
gauged minimal anomaly-free model is shown be singular,
as the kinetic terms of the quasi-Goldstone fermions vanish
in the vacuum. Gauging only the linear isometry group 
$SU(5) \times U(1)$, or one of its subgroups, can give 
a physically well-behaved theory. With gauged $U(1)$ this 
requires the Fayet-Iliopoulos term to take values in a 
specific limited range. 
\nl 
} 
 
\np 

\pagestyle{plain}
\pagenumbering{arabic}
%
%
\section{Introduction \label{Intro}}

\nit 
$N = 1$ supersymmetry in 4-$D$ space-time is likely to be a 
major ingredient of effective field theories of fundamental 
interactions at energies below or near the Planck scale. When 
restricted to theories at most quadratic in field gradients, 
the full spectrum of models is characterized by the field 
content (e.g., the spectrum of scalar and vector multiplets), 
which includes fixing the group of local gauge symmetries and 
their representations; and furthermore by the choice of three 
functions of the scalar multiplets, the K\"{a}hler potential, 
the superpotential and the holomorphic kinetic  functions of 
the gauge fields. Even with such a restricted set of choices, 
a surprisingly rich variety of structures can be realized; as 
yet their classification is far from complete. 

In this paper we discuss supersymmetric $\gs$-models on 
K\"{a}hler coset spaces $G/H$. Such models have been studied 
previously by various groups of authors \ct{1}; for some 
reviews see ref.\ct{2,2a,BL}. They have been considered in the 
context of non-standard superunification models, of effective 
low-energy models for gauge theories in the strong-coupling 
limit, or as models for string-inspired low-energy phenomenology. 
Supersymmetric $\gs$-models of various kinds are also part 
of supergravity theories. In $N = 1$ supergravity K\"{a}hler 
type models are of interest because they can realize many 
varieties of non-linear symmetries on chiral fermions. In 
extended supergravity $\gs$-models are a basic part of the 
theory, cf.\ the non-compact models on $SU(1,1)/U(1)$ in 
$N = 4$, and on $E_{7\,(+7)}/SU(8)$ in $N = 8$ supergravity 
in 4-$D$ space-time. 

Supersymmetry requires the target space of $N = 1$ scalar 
superfield theories in $D=4$ space-time to be a complex 
manifold of the K\"{a}hler type \ct{BZ;FAG}. For a 
coset model to be K\"{a}hler imposes special conditions 
on the groups $G$ and $H$; in particular, the stability 
group $H$ always factorizes so as to possess at least one 
commuting $U(1)$ subgroup. The more special {\em symmetric 
cosets} with such a structure include the Grassmannian 
models on $U(N,M)/U(N) \times U(M)$, the orthogonal 
unitary coset models on $SO(2N)/U(N)$, as well as models 
on exceptional cosets like $E_6/SO(10) \times U(1)$. 
A non-symmetric model of phenomenological interest is for 
example the supersymmetric version of $E_8/SO(10) \times 
SU(3) \times U(1)$. 

By themselves, homogeneous supersymmetric coset-models are 
known to be inconsistent quantum field theories, because 
of the appearance of anomalies in the holonomy group 
\ct{3}. These can not be compensated by Wess-Zumino type 
modifications \ct{BL}. A particular solution  to this 
problem has been proposed in \ct{WL,GS}, involving a 
procedure known as Goldstone boson doubling. This procedure 
takes a complexification of the broken isometry group as the  
starting point for the construction of field-theory models.  
Alternatively, in \ct{jw1} it was proposed to cancel such  
anomalies by coupling additional (chiral) matter superfields  
in non-trivial representations of the isometry group of the  
$\gs$-model. More details of this procedure have since been  
worked out in \ct{GNvH1,GNvH2,GN}. We wish to stress, that
although the two approaches have rather different starting 
points, they are not mutually exclusive. 

Continuing our line of investigation, in this paper we 
describe several new results. First, we perform a quite 
general analysis of the global aspects of the geometry and 
isometries of K\"{a}hler-type coset manifolds. From the 
results we derive a better and more detailed understanding 
of the consistency conditions on the bundles which can be 
constructed over such manifolds. As these bundles can be 
interpreted as target spaces of fields coupled to the 
$\gs$-model, the consistency conditions have direct 
implications for the existence of interacting field theories 
constructed on the basis of pure coset models. We apply the 
results of this analysis to the particular case of symmetric 
orthogonal unitary cosets on $SO(2N)/U(N)$. We show that 
only a finite number of these models can be consistent when 
coupled to matter superfields with $U(N)$ quantum numbers 
reflecting spinorial representations of $SO(2N)$. Among these 
are in particular the ones based on $SO(10)/SU(5)\times U(1)$, 
with matter in representations descending from the $\underline 
{16}$ of $SO(10)$, which are interesting for phenomenological 
applications. 

This paper is structured as follows. In section \ref{s2} 
we review some basic aspects of K\"{a}hler geometry and its 
role in supersymmetric scalar field theories in $D = 4$ 
space-time. We discuss the symmetries of these models 
in the geometric language of Killing vectors (generating 
isometries), which represent infinitesimal, but generally 
non-linear, transformations. A procedure for coupling chiral 
superfields in other representations of the isometry group, 
first described in \ct{CL,jw1,GNvH1}, is reviewed emphasizing 
its role in anomaly-cancelation. In section \ref{s3} we 
present the construction of non-linear realizations of the 
$SL(N+M;{\Cplx})$ starting from the approach of ref.\ 
\ct{BKMU}. By imposing certain constraints on the group 
elements one obtains non-linear representations of various 
classical groups, like $SU(N+M)$, $SO(2N)$, or $USp(2N)$ 
and their non-compact relatives, in finite form. We discuss 
the realization of the non-linear transformations on 
various types of bundles over the manifold, and examine 
the consistency conditions to be satisfied. In the context 
of field theory, this results in the quantization of $U(1)$ 
charges for matter fields coupled to the $\gs$-model. 
In section \ref{s4} we turn specifically to non-linear 
realizations of $SO(2N)$. The bundles of interest for 
supersymmetric field theory applications are constructed. 
We use the conditions for existence of these bundles to
examine the possibility of cancelling anomalies in section 
\ref{s5}. We identify $U(N)$-bundles over the $SO(2N)/U(N)$ 
cosets which together build a spinor representation of 
$SO(2N)$. It is shown that only a finite number of spinor 
models can be made anomaly-free. We finish in section 
\ref{s6} by discussing a number of physical aspects of the 
model on $SO(10)/U(5)$, like internal and supersymmetry 
breaking. A rather surprising result, which generalizes to 
other coset models, is that the model with fully gauged 
$SO(10)$ is singular: the kinetic terms of the Goldstone 
superfields vanish in the vacuum. Gauging the linear subgroup 
$U(5)$ can give consistent models, but only in a range of 
non-zero values of the Fayet-Iliopoulos term. Finally, 
the appendices describe some mathematical details  of our 
constructions. 

%
%

\section{$\gs$-models on K\"{a}hler manifolds 
\label{s2}} 

The kinetic terms of complex scalar fields $(z^{\ga},z^{\uga})$ 
on K\"{a}hler manifolds are of the form 
\begin{equation} 
{\cL}_{kin}\, =\, -\ \sqrt{-g}\, g^{\mu\nu}\, G_{\ga \uga}(z,
 \bar{z})\, \der_{\mu} z^{\ga}\, \der_{\nu} \bar{z}^{\,\uga}, 
\label{1.1} 
\end{equation} 
with the spacetime metric $g_{\gm\gn}$ and 
the target-space metric subject to the K\"{a}hler condition 
\begin{equation} 
G_{\ga\, \uga, \gb}\, =\, G_{\gb\, \uga, \ga}. 
\label{1.2} 
\end{equation} 
Locally on the target manifold the condition is satisfied by 
deriving the metric from a K\"{a}hler potential: 
\begin{equation} 
G_{\ga\, \uga}\, =\, K_{,\ga\, \uga}. 
\label{1.3}
\end{equation} 
In terms of the K\"{a}hler two-form 
\begin{equation} 
\omega(K)\, =\,-i\, K_{,\ga\uga}\, d\bar{z}^{\uga} \wedge dz^{\ga}
\label{1.3.1}
\end{equation} 
equation (\ref{1.2}) can be written as 
\begin{equation} 
d \omega(K)\, =\, 0. 
\label{1.3.2}
\end{equation} 
Obviously, the K\"{a}hler potential is defined by this equation 
only up to holomorphic terms: 
\begin{equation} 
\tK(z, \bar{z})\, =\, K(z, \bar{z})\, +\, F(z)\, +\, 
 \bar{F}(\bar{z}). 
\label{1.4} 
\end{equation} 
As a result, if two complex local co-ordinate charts 
$\{z_i\}$ and $\{z_j\}$ have non-empty overlap, the 
K\"{a}hler potentials in the charts are generally related by 
\begin{equation} 
K_i(z_i,\bar{z}_i)\, =\, K_j(z_j,\bar{z}_j)\, +\, 
 F_{(ij)}(z_j)\, +\, \bar{F}_{(ij)}(\bar{z}_j). 
\label{1.5}
\end{equation} 
In this paper we focus on the class of K\"{a}hler manifolds 
formed by coset spaces $G/H$. Such cosets are K\"{a}hler 
manifolds if $H$ is the centralizer of a torus in $G$; that 
is, the stability group $H$ contains one or more $U(1)$ 
subgroups commuting with the rest of $H$. A general procedure 
for constructing a K\"{a}hler potential for these coset 
manifolds was developed by the authors of ref.\ \ct{BKMU}.
It is discussed in some detail in the context of our 
applications below. 

Scalar lagrangeans of the type (\ref{1.1}) can be extended 
to incorporate $N = 1$ Poincar\'{e} supersymmetry, by taking 
the complex fields $z^{\ga}$ to be the scalar components 
of chiral superfields $\gF^{\ga}$; we denote its chiral fermion 
components by $\psi_L^{\ga}$. In Minkowski space, the component 
lagrangean is \ct{BZ;FAG}
\begin{eqnarray} 
\cL_{chiral} & = & \dsp{ \int d^4\theta\, K(\gF,\bar{\gF}) }
   \label{1.7} \\ 
 & & \nonumber \\ 
  & = & \dsp{ -\, G_{\ga \uga}(z, \bar{z})\, \left[ 
 \der^{\mu} z^{\ga}\, \der_{\mu} \bar{z}^{\,\uga} + 
 \bar{\psi}_L^{\uga} \stackrel{\leftrightarrow}{\sDer} 
 \psi_L^{\ga} \right]\, +\, \frac{1}{4}\, R_{\ga\uga\gb\ugb}\, 
 \bar{\psi}_L^{\uga} \gg^{\mu} \psi_L^{\ga}\, 
 \bar{\psi}_L^{\ugb} \gg_{\mu} \psi_L^{\gb}, } \nonumber
\end{eqnarray} 
where the covariant derivative and curvature tensor are 
those of the K\"{a}hler manifold. 

In general, K\"{a}hler metrics admit a set of holomorphic 
isometries $R_i^{\,\ga}(z)$, with conjugates 
$\bar{R}^{\,\uga\,}_i(\bar{z})$, satisfying the Killing 
equation 
\begin{equation} 
R_{i\,\uga, \ga}\, +\, \bar{R}_{i\,\ga, \uga}\, =\, 0. 
\label{1.8} 
\end{equation} 
These isometries define infinitesimal symmetry transformations 
on the manifold: 
\begin{equation} 
\delta z^{\ga}\, =\, z^{\prime\, \ga}\, -\, z^{\ga}\, =\, 
 \gth^i\, \delta_i z^{\ga}\, =\, \gth^i\, R_i^{\,\ga}(z), 
\label{1.9}
\end{equation} 
with $\gth^i$ the parameters of the infinitesimal 
transformations. As a result, the isometries define a Lie 
algebra with structure constants $f_{ij}^{\;\;\;k}$ via the 
Lie derivative by: 
\begin{equation} 
\lh {\cal L}_{R_i}[R_j] \rh^{\ga}\, =\, 
R_i^{\, \gb} R^{\,\ga}_{j, \gb}\, -\,  
R_j^{\, \gb} R^{\,\ga}_{i, \gb}\, =\, f_{ij}^{\;\;\;k}\, 
R_k^{\, \ga}.
\label{1.10} 
\end{equation} 
The invariance of the metric implies, that under these isometries 
the K\"{a}hler potential generally is invariant modulo holomorphic 
functions, as in eq.\ (\ref{1.4}):  
\begin{equation} 
\delta_i K\, =\, F_i(z) + \bar{F}_i(\bar{z}). 
\label{1.10.1} 
\end{equation} 
From the Lie-algebra property (\ref{1.10}) it follows that one
can choose the transformations of the functions $F_i(z)$ to 
have the property 
\begin{equation} 
\delta_i F_j\, -\, \delta_j F_i\, =\, f_{ij}^{\;\;\;k}\, F_k. 
\label{1.10.2}
\end{equation} 
Equation (\ref{1.8}) for holomorphic Killing vectors has a local 
solution in terms of a set of scalar potentials $M_i(z,\bz)$, 
transforming in the adjoint representation of the Lie-algebra 
(\ref{1.10}):
\begin{equation} 
-i M_i = K_{,\ga} R^\ga_i - F_i, 
\hspace{3em} 
\delta_i M_j\, =\, f_{ij}^{\;\;\;k} M_k.  
\label{Killing_Pot}
\end{equation}   
Supersymmetry generalizes the isometries to transformations of
the chiral superfields $\gF^{\ga}$: 
\begin{equation} 
\delta_i \gF^{\ga}\, =\, R_i^{\ga}(\gF). 
\label{1.11}
\end{equation} 
For the chiral fermions this implies the infinitesimal 
transformation rule 
\begin{equation} 
\delta_i \psi_L^{\ga}\, =\, R_{i\:\: ,\gb}^{\ga}(z)\, 
 \psi_L^{\gb}.  
\label{1.12}
\end{equation} 
In sect.\ \ref{s3} we present the finite form of the 
transformations (\ref{1.11}), of which (\ref{1.9}) and 
(\ref{1.12}) are special cases, for a large class of 
symmetric coset spaces $G/H$. 

As the chiral fermions couple to the connection and the curvature 
in ${\cL_{chiral}}$, the consistency of the quantum theory is 
generally spoiled by anomalies \ct{2}. Therefore we extend 
the model with additional chiral superfields ---generically 
called {\em matter} superfields--- on which the isometry group 
is realized, with the representations chosen to cancel the 
anomalies. 

A general procedure for matter coupling has been worked out 
in \ct{CL,jw1,GNvH1}; the generalization to supergravity was 
presented in \ct{GNvH2}. The mathematical framework used to 
construct matter representations of the isometry group of 
the K\"{a}hler manifold is the theory of complex bundles 
over K\"{a}hler manifolds. These bundles are defined locally 
on the K\"{a}hler manifold by sets of complex fields with 
specific transformation character under the isometries. 

The basic pattern is that exhibited by the transformation rule 
(\ref{1.12}) for the chiral fermions. This rule shows how a 
vector (an element of the tangent bundle) transforms under 
the isometries. Similarly, one can define a representation 
transforming as a 1-form (an element of the co-tangent bundle): 
\begin{equation} 
\delta_i v_{\ga}\, =\, - R^{\,\gb}_{i\; ,\ga}(z)\, v_{\gb}. 
\label{1.13}
\end{equation}  
More general tranformations are obtained by taking tensor 
products of the tangent or co-tangent bundles. However, for 
our applications this is not sufficient. The reason is, that 
the $U(1)$ charges of such representations are completely 
fixed in terms of the charge of the scalars $z^{\ga}$: a 
contravariant holomorphic tensor $t^{\ga_1...\ga_p}$ of 
rank $p$ carries a relative charge $p$, whereas a covariant 
holomorphic tensor $s_{\ga_1...\ga_k}$ of rank $k$ carries 
a relative $U(1)$ charge $-k$. But in actual models, if one 
requires anomaly cancellations with a phenomenologically 
interesting set of matter superfields, one usually needs 
a different assignment of $U(1)$ charges. Therefore the 
spectrum of representations must be extended with bundels 
which differ from tensor bundles by the assignment of $U(1)$ 
charges. This is achieved for instance by the introduction 
of complex line bundles \ct{GNvH1}.  

A line bundle is the target space of a single-component 
complex scalar field over the manifold. We consider line 
bundles carrying non-trivial representations of the isometry 
group; these can be defined locally on the K\"{a}hler 
manifold as complex scalar matter fields $S(x)$ coupled to 
the $\gs$-model, with the infinitesimal transformation law 
given by 
\begin{equation} 
\delta_i S\, =\, F_i(z) S. 
\label{1.14}
\end{equation} 
In the context of supersymmetric field theories such a 
representation of the iso\-metry group was introduced in 
\ct{jw1}, and subsequently considered in \ct{Elw}; it is 
a representation because of the property (\ref{1.10.2}). 
From the line-bundle $S$ one can obtain other line 
bundles with different $U(1)$ weights by taking powers: 
\begin{equation} 
A\, \equiv\, S^{\gl} \hspace{1em} \Rightarrow \hspace{1em} 
\delta_i A\, =\, \gl F_i(z) A. 
\label{1.15}
\end{equation} 
Furthermore, using the line bundle construction, one can modify 
the transformation rules of fields in tensor representations 
of the isometry group. For example, defining 
\begin{equation} 
T^{\ga_1...\ga_p}\, \equiv\, S^{\gl}\, t^{\ga_1...\ga_p}, 
\label{1.16}
\end{equation} 
the new field $T$ obeys the transformation rule 
\begin{equation} 
\delta_i T^{\ga_1...\ga_p}\, =\, \sum_{k=1}^p\, 
 R_{i\;\, ,\gb}^{\ga_k}\, T^{\ga_1..\gb..\ga_p}\, +\, 
 \gl F_i\, T^{\ga_1...\ga_p}. 
\label{1.17} 
\end{equation} 
In this way the $U(1)$ charges can be adjusted, be it subject 
to the charge quantization conditions mentioned above.  

However with the introduction of the line-bundle we still 
have not exhausted all possibilities for consistent 
non-linear realizations of symmetries over K\"{a}hler 
manifolds. Some coset spaces allow factorization of the 
goldstone-boson transformations and the K\"{a}hler 
metric. Then one can define sub-bundles of the tangent-space 
bundles, and their line-bundle extensions as well. 
A general description of matter representations that 
can be associated with coset spaces can be found in 
refs.\ \cite{2,Luty:1995ug}. Examples of 
this type of structures are presented below. 

The bundles introduced here are characterized locally on 
the K\"{a}hler manifold by their transformation properties. 
An important question is, if these definitions can be 
extended globally over the manifold. This is always possible
for tangent and co-tangent bundles. However, for line bundles
(\ref{1.14}), this requires in particular that the 
holomorphic transition functions introduced in (\ref{1.5}) 
satisfy the cocycle condition 
\begin{equation} 
F_{(ij)}\, +\, F_{(jk)}\, +\, F_{(ki)}\, =\, 2 \pi i \Intr.
\label{1.15.2}
\end{equation} 
Manifolds with this property are known as K\"{a}hler-Hodge 
manifolds \ct{KH}; their K\"{a}hler forms satisfy the 
condition 
\begin{equation} 
\int_{C_2}  \omega(K)\, =\, 2\pi \Intr, 
\label{1.15.1}
\end{equation} 
for any closed two-cycle $C_2$. 

The existence of the generalized line bundles (\ref{1.15}) 
and (\ref{1.17}) often requires the powers $\gl$ to 
satisfy certain integrality conditions: there is a minimal 
line bundle which by eq.\ (\ref{1.15.2}) is globally 
defined and single-valued, and all other line bundles 
carry integral charges w.r.t.\ the minimal line bundle.
Thus it follows that the $U(1)$ charges of fields 
transforming as line bundles are quantized \ct{GN}. These 
consistency requirements are discussed in detail in section 
\ref{SLNC} below. 

%
%
\section{Non-Linear Realization of 
$\boldsymbol{SL(N+M, \Cplx)}$ \label{s3}}
\label{SLNC} 

In this section we discuss the method developed by Bando, 
Kuramoto, Maskawa and Uehara (BKMU) \cite{BKMU} to 
obtaining non-linear transformations and \Kh\ potentials, but 
here we consider more general transformations of the complex 
coordinates and we discuss matter coupling in detail. 
The basis of our construction is to define a transformation 
rule for a complex $M\times N$-matrix $z$ under the action of 
an arbitrary element of special linear group $SL(M+N; \Cplx)$. 
It will become clear below, why we restrict ourselves to the 
special linear group. As the special linear group contains all 
(classical) Lie groups as subgroups, this construction can be 
used to obtain \Kh\ potentials for k\"{a}hlerian coset spaces 
based on these groups. It explains why the transformation 
rules for the different coset spaces based on classical groups 
are much alike. 
Given a coset the group of isometries is fixed, but we can still 
use the full $SL(M+N;\Cplx)$ transformations to determine the 
effect of coordinate redefinitions. In particular for a 
non-compact coset this allows us to interpolate between 
two seemingly different representations of its \Kh\ potential.

Let $g \in SL(M+N; \Cplx)$ be an arbitrary element of the 
special linear group and $g\inv$ its inverse, we write 
\equ{
g = \pmtrx{ \ga & \gb \\ \gg & \gd}
\qand
g\inv = \pmtrx{ \siga & \sigb \\ \sigg & \sigd},
\labl{Mtrxg}
}
where $\ga, \gb, \gg$ and $\gd$ are $M\times M$-, $M\times N$-,  
$N\times M$- 
and $N\times N$-matrices respectively.  The submatrices of the 
inverse $g\inv$ are given by
\equ{
\mtrx{
\mtrx{
\siga = (\ga - \gb \gd\inv \gg)\inv, \qquad
\sigd = (\gd - \gg \ga\inv \gb)\inv,
}
\\[2mm]
\mtrx{
\sigb = - (\ga - \gb \gd\inv \gg)\inv \gb \gd\inv = 
- \ga\inv \gb (\gd - \gg \ga\inv \gb)\inv,
\\[2mm]
\sigg= - (\gd - \gg \ga\inv \gb)\inv \gg \ga\inv = 
- \gd\inv \gg (\ga - \gb \gd\inv \gg)\inv.
}
}
}
To obtain the infinitesimal transformations, one considers 
infinitesimal deviations from the unit element of 
$SL(M+N; \Cplx)$
\equ{
g = \pmtrx{
\Id + u & y \\
x & \Id - v
}
\quad \text{and} \quad
g\inv  = \pmtrx{
\Id - u & -y \\
-x & \Id + v
},
\labl{Mtrxinfg}
}
where $u, v, x, y$ are infinitesimal submatrices and the minus 
in front of the $v$ is useful later. However in the following 
we are primarily concerned with finite transformations. 
A non-linear relalization is found by defining the matrix 
$\gx(z)$ like the BKMU-parameter by
\equ{
\gx(z) = 
\pmtrx{ \Id & 0 \\
z & \Id }
\label{DefXiSL}
}
and requiring that 
\equ{
\gx(z) \lra \gx(^g z) =  
g \gx( z) \hat h\inv (z; g),
\quad \text{with} \quad
\hat h = 
\pmtrx{ (\hat h_+)\inv & \hat h_0 \\
0 & \hat h_-}.
\labl{transXi}
}
We have written $(\hat h_+)\inv$ instead of $\hat h_+$ in 
the matrix $\hat h$ for later convenience, at this stage this 
is merely notation. We find that $z$ transforms as 
\equ{
\pset{g}{z} = \lh \gg + \gd z \rh \lh \ga + \gb z \rh\inv 
 = \lh \sigd - z \sigb \rh^{-1} \lh \sigg - z \siga \rh.
\labl{transz}
}
under the action of $g$ and the matrix $\hat h$ takes the form
\equ{
\hat h(z; g) = 
\pmtrx{ (\hat h_+)\inv & \hat h_0 \\
0 & \hat h_-} =
\pmtrx{
\ga + \gb z & \gb \\
0 & 
( \sigd - z \sigb  )\inv
}.
\labl{hath}
}
Notice that it follows from the transformation rule of $z$ that
general linear transformations have the same effect as special 
linear transformations. Under the composition of two 
transformations $g'$ and $g$ we find using \eqref{transXi} that 
the non-linear transformation \eqref{transz} respects this 
composition 
\(
^{g'}(^g z)\, =\, ^{g'g}z
\)
and furthermore we find that
\equ{
\hat h_-(z; g' g) = \hat h_-(^g z; g')\hat h_-(z; g)
\qand
\hat h_+(z; g' g) = \hat h_+(z; g)\hat h_+(^g z; g').
\labl{comph}
}
In the following we employ two projector operators $\get_\pm$ 
defined by 
\equ{
\get_+ = \pmtrx{ \Id & 0 \\ 0 & 0 }
\quad \text{and} \quad
\get_- = \pmtrx{ 0 & 0 \\ 0 & \Id }.
\labl{projpm}
}
These definitions allow us to write 
\(
(\hat h_+)\inv \simeq \hat h \get_+ = \get_+ \hat h \get _+
\) and
\(
\hat h_- \simeq \get_- \hat h = \get_- \hat h \get _-
\), where the symbol $\simeq$ denotes equality of the 
left-hand side as the unique non-vanishing submatrix
of the right-hand side. 

Now let $\fJ \in SL(M+N; \Cplx)$ be a fixed matrix; its 
properties we develop along the way. We define the 
$M\times M$-matrix function of $z, \bz$ by 
\equ{
\tgch\inv_\fJ(z, \bz) 
\equiv \get_+ \gx^\dag(\bz) \fJ \gx(z) \get_+ 
= A + B z + \bz C + \bz D z
\labl{ChiI}
}
and obtain the transformation property
\equ{
\tgch\inv_\fJ(z, \bz) \lra 
\tgch\inv_\fJ(^g z, ^g \bz) = 
\hat h_+^\dag(\bz; g) 
 \tgch\inv_{g^\dag \fJ g}(z, \bz)
\hat h_+ (z; g).
\labl{transChiI}
}
Define the subgroup $SL_\fJ(M+N;\Cplx)$ consisting of elements 
$g \in SL(M+N; \Cplx)$ that leave $\fJ$ invariant 
\equ{
g^\dag \fJ g = \fJ \equiv
\pmtrx{ A & B \\ C & D }, \qquad
\fJ\inv \equiv
\pmtrx{ \siA & \siB \\ \siC & \siD }.
\labl{I-SL}
}
Hence if $g \in SL_\fJ(M+N; \Cplx)$, the function 
\equ{
K_\fJ (z, \bz) = \ln \det \tgch\inv_\fJ(z, \bz)
\labl{KI}
}
transforms as a \Kh\ potential
\equ{
K_\fJ (z, \bz) \lra K_\fJ (^g z, ^g \bz) = 
K_\fJ (z, \bz)  + F(z; g) + \bar F (\bz; g),
\labl{transKI}
}
with 
\equ{
F(z; g) = \ln \det \hat h_+(z; g),
\qquad
\bF(\bz; g) = \ln \det \hat h^\dag_+(\bz; g^\dag).
\labl{defF}
}
If we want to interpret $K_\fJ$ as a \Kh\ potential, 
$K_\fJ$ has to be a real function
\(
K_\fJ(z, \bz) = \lh K_\fJ (z, \bz) \rh^\dag.
\)
This only happens iff $\fJ$ is Hermitean 
\(
\fJ^\dag = \fJ.
\)
The composition rule for $F$ follows directly from 
eq.\ \eqref{comph}
\equ{
F(z; g' g) = F(z; g) + F(^g z; g').
\labl{compF}
}
We define a real finite Killing pre-potential $\cM(\bz, z; g')$ 
by
\equ{
2i\, \cM(\bz, z; g') = 
K(z, \pset{g'}\bz) - K(\pset{g'}z, \bz) + 
F(z; g') - \bF(\bz; g').
\labl{defM}
}
It is a function of the group element $g'$, of which the 
infinitesimal (linearized) form reproduces the standard 
Killing potentials (\ref{Killing_Pot}). Using the 
transformation property of the \Kh\ potential \eqref{transKI} 
together with the composition property \eqref{compF} of $F$, 
it follows that $\cM(\bz, z; g')$ transforms in the adjoint 
representation 
\equ{
\cM(\pset{g}z, \pset{g}\bz; g') = 
\cM(z, \bz; g\inv g' g).
\labl{transM}
}
Again, inserting a group element close to the identity, 
we obtain for the Killing potentials the infinitesimal 
transformation rule (\ref{Killing_Pot}). 

The metric associated with $K_\fJ$ can be written as 
 \equ{
G_{\uga \ga} d \bz^\uga d z^\ga = 
\tr \left[
\tgch\lw\fJ d \bz 
\gch\lw{\fJ}
d z
\right],
\labl{Imetric}
}
where we define $\gch\lw{\fJ}$  in analogy of 
$\tgch\lw\fJ$ in \eqref{ChiI}
\equ{
\gch\inv_{\fJ}(z, \bz) 
\equiv \get_- \lh \gx^\dag(\bz) \fJ \gx(z) \rh\inv \get_-
= \siD - \siC \bz - z \siB  + z \siA \bz.
\labl{tChiI}
}
This can be shown either by a direct calculation of the metric 
in the standard way as the second mixed derivative or by first 
proving this for a block-diagonal $\fJ$ and showing that 
the diagonalization procedure has no effect on the metric
(\ref{Imetric}). This is easy as under the action of 
$g \in SL_\fJ(M, N)$ the differential $dz$ transforms as
\equ{
d z \lra \pset{g}{(d z)}  = 
\hat h_-(z;g) dz \hat h_+ (z; g) = 
( \sigd - z \sigb )\inv d z 
(  \ga + \gb  z )\inv,
\labl{transdz}  
}
and $\tgch\lw\fJ, \gch\lw\fJ$ transform as eq.\ 
\eqref{transChiI} and as  
\equ{
\gch\inv_{\fJ}(z, \bz) \lra 
\gch\inv_\fJ(^g z, ^g \bz) = 
\hat h_-(z; g) 
 \gch\inv_{\fJ}(z, \bz)
 \hat h_-^\dag (\bz; g).
\labl{transtChiI}
}
Hence it follows that \eqref{Imetric} is invariant.

Until this point the matrix $\fJ$ used in the definitions 
\eqref{ChiI} and \eqref{tChiI} of $\tgch\lw\fJ$ and $\gch\lw\fJ$ 
can be any Hermitean matrix of $SL(M+N; \Cplx)$. However if we 
want to use the invariants \eqref{invsLR} as \Kh\ potentials for 
supersymmetric model building, the resulting kinetic terms have 
to be positive definite. By going to the unitary gauge 
($\bz = z^T = 0$), we infer that both $A$ and $\siD$ have to be 
sign definite. (Of course an overall sign can be compensated by 
an appropriate minus sign.) On the other hand using a unitary 
transformation, we can diagonalize $\fJ$ with real eigenvalues 
$\gl_i$. If this is followed by an appropriate scale 
transformation of the coordinates and possibly some relabeling,  
we bring the matrix $\fJ$ into the canonical form 
\equ{
\fJ = \pmtrx{
\Id & ~~0 \\
0 & \get \Id
}, 
\qquad \get = \pm 1.
\labl{SUform}
}
This shows that we can restrict $SL_\fJ(M+N; \Cplx)$ 
to  $SU_\get(M,N)$ when we want to study the isometries of the 
metrics $\tgch_\fJ$, $\gch_\fJ$ or the \Kh\ potential $K_\fJ$. 
Here $\get = +1$ refers to the compact special unitary group 
$SU(M + N)$, while $\get = -1$ refers to the non-compact version. 
We assume from now on that we have chosen this canonical form of 
$\fJ$ and consider $SU_\get(M, N)$ only. Notice that by putting 
further restrictions on the group elements $g$ we can reduce the 
isometry group to a subgroup of $SU(M,N)$, such as $SO(2N)$ or
$USp(N)$. The form of the metrics and \Kh\ potential does not 
change under this; they always take the form
\equ{
\mtrx{
\tgch_\get (z, \bz) =  \lh \Id + \get \bz z \rh\inv, \quad 
\gch_\get (z, \bz) = \lh \Id + \get z \bz \rh\inv,
\\[2mm]
K_\get (z, \bz) = \get \ln \det \tgch\inv_\get = \get \ln \det 
\gch\inv_\get
}
\label{CanonicalKh}
}
in the canonical basis. However this leads to restrictions on 
the coordinates $z$ as we see later: that is the coordinates $z$ 
parameterize a submanifold of $SU_\get(M,N)/S[U(M)\times U(N)]$. 
Even though the $SL(M+N; \Cplx)$ group is not the isometry group, 
it is still worthwhile to know its action on the fields, as it 
can be used to describe field redefinitions. 

We give an example of this now. In the previous analysis we used 
that we can set $B$ and $C$ in the matrix $\fJ$ to zero by a 
unitary transformation. Sometimes we can also do the opposite: 
set $A$ and $D$ to zero. To analyze the situation we start 
with $\fJ$ in the canonical form and perform an arbitrary 
transformation $g$ of $SL(M+N;\Cplx)$ on it
\equ{
g^\dag \pmtrx{ \Id & 0 \\ 0 & \get \Id } g = 
\pmtrx{
\bga \ga + \get \, \bgg \gg & \bga \gb + \get \, \bgg \gd 
\\[2mm]
\bgb \ga + \get \, \bgd \gg & \bgb \gb + \get \, \bgd \gd 
}.
}
So to remove the $A$ and $D$ entries of this matrix we need 
to have that 
\equ{
\bga \ga + \get \, \bgg \gg = 0
\quad \text{and} \quad 
\bgb \gb + \get \, \bgd \gd = 0.
}
Notice that there is no solution $g \in SL(M+N; \Cplx)$ 
of these equations when $\get = 1$. On the other hand in the 
case $\get = -1$ with $M = N$ we can use 
\equ{
g = \frac i{\sqrt 2} \pmtrx{
- \Id & - \Id \\
- \Id & ~~\Id
}
} 
to bring $\fJ$ into the form
\equ{
\fJ = \pmtrx{ 0 & \Id \\ \Id & 0 }.
}
Using this matrix $\fJ$ we obtain a \Kh\ potential
\equ{
K_{no-sc} = \ln \det ( z + \bz )
}
of the no-scale type \cite{Lahanas:1987uc}. The low energy 
effective actions for the moduli sectors of string theory often 
take this form, see for example 
\cite{GSW-II,Polchinski,Quevedo:1996sv}. \nl

\noindent
Matter coupling is the next topic we discuss. As we want 
to interpret $SU_\get(M,N)$ as the symmetry group of the models 
we construct, this implies that all matter representations 
should be well defined representations of $SU_\get(M, N)$. 
To obtain a section of the tangent bundle, we define the 
transformation of the tangent space vector $T$ in analogy of 
\eqref{transdz} by
\equ{
\pset{g}{T} = 
\hat h_-(z;g) T \hat h_+ (z; g) = 
(\sigd - z \sigb)\inv T (\ga + \gb z)\inv.
}
A section $C$ of the cotangent bundle transforms as
\equ{
\pset{g}{C} = 
\lh \hat h_+(z; g) \rh\inv C \lh \hat h_-(z; g)\rh\inv = 
 (\ga + \gb z) C (\sigd - z \sigb) .
}
When we take $g \in SU_\get (M,N)$, we obtain the following 
invariants for the sections of the tangent and cotangent bundles
\equ{
\tr \left[
\tgch\lw\fJ \bar T
\gch\lw{\fJ} T
\right]
\qand
\tr \left[
(\gch\lw\fJ)\inv \bar C
(\tgch\lw{\fJ})\inv C
\right].
}
Next we construct subbundles of the tangent bundle. To do this 
we notice that the transformation rule \eqref{transdz} for the 
differential $d z$ factorizes \cite{GN,Luty:1995ug}. Using this 
we define the sections $L$ and $R$ by  the transformation rules 
\equ{
\pset{g}{L}  = \hat h_-(z;g) L = (\sigd - z \sigb)\inv L
\quad \text{and} \quad 
\pset{g}{R}  = R \hat h_+(z;g)  = R (\ga + \gb z)\inv.
\labl{transsLR}
}
To show that these transformations do indeed define consistent 
bundles we proceed as follows. All manifolds we consider here 
are submanifolds of Grassmannian coset-space 
$SU_\get(M,N)/S[U(M)\times U(N)]$. As this is a homogeneous 
space we can reach any point on it by a transformation using a 
group element $g \in SU_\get(M,N)$. Therefore we can describe 
all coordinate transformations as actions of elements of 
$SU_\get(M,N)$, and the holomorphic transition functions on 
overlapping complex co-ordinate charts for the bundle of which 
$L$ is a section are given by elements $\hat  h_-(z; g)$. The 
global consistency conditions for this bundle, mentioned in 
sect.\ (\ref{s2}), then take the form 
\equ{
\mtrx{
\hat  h_-(z; e) = \Id, \quad
\hat h_-(^g z; g\inv) = \hat h_-(z; g)\inv 
\\[2mm]
\hat h_-(^{g_2 g_1} z; g_3) \hat h_-(^{g_1}z; g_2) 
\hat h_-(z; g_1) = \Id,
}
\label{CocycleSU}
}
when $g_1 g_2 g_3 = e$, being $e$ the $SU_\get(M,N)$ identity. 
The composition property \eqref{comph} of two group elements 
show that these conditions are satisfied. Using the metric of 
the tangent bundle \eqref{Imetric}, which factorizes as well, 
we obtain the following $SU_\get(M, N)$-invariants 
\equ{
\bL \gch\lw{\fJ} L  
\quad \text{and} \quad 
R \tgch\lw\fJ \bR.
\labl{invsLR}
}
We will discuss tensor products of these types of matter 
representations extensively when we consider matter coupling to 
$SO(2N)/U(N)$.

Until this point our discussion was general, in the sense that 
we only demanded that we construct isometries of the metrics 
$\tgch\lw\fJ$ and $\gch\lw\fJ$ without any reference to a 
particular coset space. We saw that we only obtain isometries 
of these metrics if we restrict the transformations to be 
unitary $g \in SU_\get(M,N)$. It is now easy to describe 
non-linear realizations of (classic) groups, that are subgroups 
of $SU_\get(M, N)$. For this we only have to describe what the 
group and algebra of the groups look when embedded in the 
unitary group $SU_\get(M, N)$. We have summarized our results 
in table \ref{EmbeddingSU}. We describe the ingredients of this 
table which are partly taken from ref.\ \cite{Gilmore:1974}. A 
complete classification of \Kh\ cosets can be found in ref.\ 
\cite{Helgason:1978}. A discussion on $SO(2N)/U(N)$, 
$Sp(2N)/U(N)$ cosets can also be found in refs.\ 
\cite{Ong:1983uj,Gomes:1984gu,Delduc:1985sz}. 

The classic groups are either real or complex groups that 
satisfy certain Hermitean conjugation and transposition 
properties
\equ{
g^\dag \fJ g = \fJ
\qand
g^T \fK\, g = \fK
\label{GroupDefs}
}
where $\fJ$ and $\fK$ are fixed matrices. We discriminate 
between the unitary ($SU$), orthogonal ($SO$), symplectic 
($Sp$) and unitary symplectic ($USp$) groups. Furthermore, 
with $\get = \pm 1$ we make a distinction between compact 
($\get = 1$) and non-compact ($\get = -1$) groups. We require 
the maximal subgroups $\text{H}$  of these groups to have a 
compact $U(1)$-factor. For example, we do not consider the 
non-compact $SO(N,N)$ here, as the non-compact abelian $SO(1,1)$ 
subgroup corresponds to Lorentz transformations that are not 
bounded. A compact $U(1)$-factor is needed to ensure that the 
resulting coset-space is \Kh; because of its importance we give 
the $U(1)$ embedding explicitly. For the real groups $SO(2N)$ 
and $Sp(2N)$ the $U(1)$ is not realized in a diagonal way. By 
making a similarity transformation 
\equ{
g_D = V g V^\dag, \quad g = V^\dag g_D V
\qand
V = \frac 1{\sqrt 2} \pmtrx{ \Id & i \\ i & \Id },
}
using the unitary matrix $V$, the $U(1)$ is turned into a 
diagonal form. Here the subscript $D$ is used to indicate that 
$g_D$, for example, is considered in the basis where the $U(1)$ 
is diagonal. As $V$ is unitary, $g_D$ has the same unitary 
properties as $g$. However the transposition properties may 
change
\equ{
g_D^T \fK_D g_D = \fK_D = \lh V^\dag \rh^T \fK\, V^\dag.
}
For this it is crucial that we have embedded the real groups 
$Sp(2N)$ and $SO(2N)$ in the special unitary group $SU(N,N)$ 
and $SU(2N)$ respectively; else the multiplication with $i$ 
has no meaning. In the remainder we work in the basis where 
the $U(1)$-factor is diagonal. We can now represent any element 
of any of these groups as a unitary matrix 
\(
g_D = e^{a_D},
\)
that is obtained by exponentiating an anti-Hermitean 
algebra element $a_D$. The group definition properties 
\eqref{GroupDefs} can be written down for the algebra elements 
$a_D$ as well
\equ{
a_D^\dag = - \fJ a_D \fJ\inv 
\qand
a_D^T = - \fK_D a_D \fK_D\inv.
\label{AlgebraDefs} 
}
Using these properties it is possible to give a unique 
representation of the algebra elements $a_D$. For the different 
groups we give this representation in the row of $g_D$ in 
table \ref{EmbeddingSU}. Notice that algebra elements of $Sp(2N)$ 
and $USp(N,N)$ have the same representation in the basis where 
the $U(1)$ is diagonal; therefore their corresponding cosets are 
isomorphic. From this representation of the algebra, it is easy 
to see what the restrictions are on the coset coordinates $z$ 
for the different coordinates. For the non-compact coset, the 
coordinates in addition satisfy $ \tr (z \bz) < 1$ for the 
\Kh\ potential and metrics in \eqref{CanonicalKh} to be well 
defined. Notice that for the $USp, Sp$ and $SO$ cosets the 
submetrics $\tgch\lw\get, \gch\lw\get$ are each others transposed
\(
\tgch\lw\get = \gch_\get^T.
\)

\begin{table}
{\scriptsize
\cntr{\(
\arry{| c | c  c  c  c  |}{1.2}{
\hline
\text{Group G}  & 
SU_\get(M,N) & USp_\get(N, N) & Sp(2N) & SO(2N) 
\\[2mm]
~\get = &  
\pm 1 & \pm 1 &  -1 & 1
\\[2mm] 
\text{Compact subgroup} ~ H &
S[U(M) \times U(N)] & 
U(N) & U(N) & U(N)
\\[2mm]
g \in  &
SL(M+N; \Cplx) & SL(2N; \Cplx) &
SL(2N; \Real) & SL(2N; \Real)
\\[2mm]
g^\dag \fJ g = \fJ =  & 
\dsp{\pmtrx{1 & 0 \\ 0 & \get 1}} & \pmtrx{1 & 0 \\ 0 & \get 1} & 
- & - 
\\[5mm]
g^T \fK\, g = \fK =  & 
- & \pmtrx{ 0 & 1 \\ -1 & 0 } & 
\pmtrx{ 0 & 1 \\ -1 & 0 } &  \pmtrx{1 & 0 \\ 0 & 1 }
\\[5mm]
U(1)~\text{embedding} & 
\pmtrx{ e^{i\frac{N \gth}P}  & 0 \\ 
0 & e^{-i\frac{M\gth}P}} 
&
\pmtrx{ e^{i{\gth}}  & 0 \\ 
0 & e^{-i{\gth}}} 
&
\pmtrx{ \cos \gth  & \sin \gth  \\ 
- \sin \gth  & \cos \gth } 
&
\pmtrx{ \cos \gth  & \sin \gth \ \\ 
- \sin \gth  & \cos \gth }
\\[5mm]
g_D^T \fK_D g_D = \fK_D =  & 
- & 
 \pmtrx{ 0 & 1 \\ -1 & 0 }
 & \pmtrx{ 0 & 1 \\ -1 & 0 } & 
-i  \pmtrx{ 0 & 1 \\ 1 & 0 }
\\[5mm]
g_D = e^{a_D},~ a_D = & 
\pmtrx{ u &  - \get x^\dag \\ x & - v} &
\pmtrx{ u &  - \get x^\dag \\ x & -u^T} &
\pmtrx{ u &  x^\dag \\ x & -u^T} &
\pmtrx{ u &  - x^\dag  \\ x& -u^T}
\\[5mm]
\text{Restrictions} & 
u^\dag = -u,~ v^\dag = -v & u^\dag = -u,~ x^T = x & 
u^\dag = -u,~ x^T = x & u^\dag = -u,~ x^T = - x 
\\
& \tr\,  u = \tr\,  v & & &
\\[2mm]
z \in G/H, z^{ij}\in \Cplx &
- & z^T = z & z^T = z & z^T = - z
\\ \hline
}
\)}}
\capt{ 
This table gives an 
overview of the (classical) Lie-groups that can be embedded 
into $SU_\get(M,N)$. With the parameter $\get$ we distinguish 
between compact ($\get = 1$) and non-compact ($\get = -1$) 
groups. For these Lie-groups the non-linear $SL(M+N; \Cplx)$
transformation rules given in this section can be used 
directly. $P = \text{gcd}(M,N)$ is defined a the 
greatest-common-divisor of $M$ and $N$. When the $U(1)$ is 
not diagonal, we have to perform a special unitary 
transformation to make it diagonal; when doing so the 
transposition properties may change. The Hermitean form of 
an element of the algebra after possible diagonalization is 
denoted by $a_D$. 
The matrices $u$, $v$, $x$ are all taken to be complex, their 
additional properties are given in second last row in the table.
The last row summarizes the symmetry properties of the coset 
coordinate matrices.
}
\label{EmbeddingSU}
\end{table} 

We now turn to the construction of the minimal complex line 
bundles. In this discussion we have to make a distinction 
between the different coset spaces as becomes clear below.  
Our discussion here is complementary to ref.\ \cite{GN} where 
general results have been presented, which we apply here
to the particular cosets discussed in this section. 

A section $S$ of a complex line bundle can be defined to 
transform as 
\equ{
{}^g S = \det \hat h_+(z; g)  S =  \det \hat h_-(z; g) S.
}
Here we have used that $\det \hat h_+ = \det \hat h_-$, 
which follows from \eqref{transXi} since $g \in SU_\get(M,N)$. 
The consistency of this complex line bundle follows 
directly from \eqref{CocycleSU} and the properties of the 
determinant. To show that we have obtained the minimal 
line bundle in the compact situation, 
we have to show that the integral over the corresponding \Kh\ form 
\equ{
\int_{C_2} \go(K) = 2\gp \, n, 
\qquad \text{with} \qquad n = \pm 1, 
\label{CocycleInt}
}
when integrated over a generating two-cycle $C_2$. 

We first turn to a Grassmannian coset $SU(M+N)/S[U(M) \times 
U(N)]$. Let $v$ be the complex coordinate of the stereographic 
projection of the complex projective line $\Cplx P^1$. 
We define a generating two-cycle by the embedding of $\Cplx P^1$ 
in the coset by taking all the coordinates $z^{ij}$ zero except 
for one which is equal to $v$. Now since the \Kh\ potential 
restricted to this embedding to $\Cplx P^1$ is given by 
\(
K(z, \bz)|^{}_{\Cplx P^1} = \ln(1 + \bv v) 
= K^{}_{\Cplx P^1}(v, \bv) ,
\)
which is the \Kh\ potential of $\Cplx P^1$ that 
satisfies $\int_{\Cplx P^1} \go(K_{\Cplx P^1}) = 2 \gp$, 
it follows that we have obtained a minimal line bundle.  
Next we discuss the compact $USp(2N)/U(N)$ and 
$SO(2N)/U(N)$ coset spaces. The coordinates of these 
spaces  satisfy 
$z^T = z$ resp. $z^T = - z$, see table \ref{EmbeddingSU}.
Therefore it is not possible to set all coordinates to 
zero except for one, except when one takes this 
symmetrization into account: 
\(
z = \pm z^T = v.
\)
Hence we find in these cases that 
\(
K(z, \bz)|^{}_{\Cplx P^1} = 2 \ln(1 + \bv v) 
= 2K^{}_{\Cplx P^1}(v, \bv) ,
\)
so that $n = 2$ in eq.\ \eqref{CocycleInt}. 
This implies that the section $S$ is the square of the minimal 
line bundle. Since the \Kh\ potential of a coset is unique up to 
a normalization factor, it follows that a section of a minimal 
line bundle over $USp(2N)/U(N)$ or $SO(2N)/U(N)$ is 
given by
\equ{
{}^g S = \lh \det \hat h_+(z; g)\rh^{\half}  S =  
\lh \det \hat h_-(z; g) \rh^{\half} S.
\label{LineUSpSO}
}
The only possible ambiguity for a global definition resides in 
the square root, it can be removed by using the BKMU-construction 
with the representation with highest weight that has all its Dynkin 
label zero except for the $N$th one \cite{GN}. 

We now determine the relative charges of the coordinates $z$, 
the matter fields $L$ and $R$, and the sections of the minimal 
line bundles, using the $U(1)$ embedding presented in table 
\ref{EmbeddingSU}. We first discuss the Grassmannian cosets 
and after  that the cosets $USp(2N)/U(N)$ and $SO(2N)/U(N)$. The 
$U(1)$-factor in $SU_\get(M, N)$, that is not in  $SU(M)\times 
SU(N)$, can be given by 
\equ{
u_\gth = \pmtrx{
e^{-iN\gth/P} \Id & 0 \\0 & e^{iM\gth/P} \Id },
}
where $P =\text{gcd}(M,N)$ 
is the greatest-common-divisor of $M$ and $N$. The 
smallest period of this $U(1)$ is $\gth = 2\pi$, since the integers 
$N/P$ and $M/P$ are relatively prime by construction. 
It follows that the coordinates $z$ have charge $(M+N)/P$ in this 
normalization. For the matter couplings $L$ and $R$ we find the 
charges $N/P$ resp. $M/P$. The section of the minimal line 
bundle has a charge $MN/P$.
For the cosets $USp(2N)/U(N)$ and $SO(2N)/U(N)$ we always 
obtain integer charges when we choose a slightly different 
normalization for $u_\gth$ given by 
\equ{
u_\gth = \pmtrx{
e^{-i2\gth} \Id & 0 \\0 & e^{i2 \gth} \Id }.
}
In this case $L$ and $R$ have the same charge $2$ and the 
section of the minimal line bundle has charge $N$, 
while the charge of the coordinates is $4$. 

%
%
\section{$\boldsymbol{SO(2N)/U(N)}$ coset models } 
\label{s4}

We discuss supersymmetric models build using the \Kh\ geometry of 
the coset space $SO(2N)/U(N)$. For this we first discuss the 
decomposition of the $SO(2N)$ algebra into $U(N)$ 
representations and the vector representation of $SO(2N)$. 
We  discuss the construction of the \Kh\ potential using the 
the general BKMU method and related that to our discussion on 
special linear transformations of section \ref{SLNC}. 
Next we discuss matter representations that can be coupled the 
supersymmetric $\gs$-model of the coset in a consistent way, 
heavily relying on the non-linear transformation discussed 
in section \ref{SLNC}. For applications to a chiral spinor 
representation of $SO(2N)$ later in this article we confine 
ourselves to the construction of completely anti-symmetric 
tensor representations with an arbitrary rescaling charge. 
We discuss their transformation properties and their invariant 
\Kh\ potentials that can be used in supersymmetric model building.  
Some relevant results and conventions have been collected in the 
appendices. 

%
%
\subsection{$\boldsymbol{SO(2N)}$ Algebra in a $\boldsymbol{U(N)}$ 
            basis} \label{SOinU}

In this section we discuss how the algebra of $SO(2N)$ can 
be decomposed  into $SU(N)\times U(1)$ representations. 
We split the $SO(2N)$ generators $M_{ab}$ into $SU(N)$ 
generators $T^i\,_j$, a $U(1)$-factor generator Y and 
broken generators $X^{ij}, \bX_{ij}$ which are anti-symmetric 
tensors of $SU(N)$. We first discuss the embedding of $U(N)$ 
in $SO(2N)$, then we discuss the vector representation; the 
spinor representation is discussed in appendix A. 

The $\smash{2N(2N-1)/2}$ anti-Hermitean generators $M_{ab} 
= - M_{ba}$ of $SO(2N)$ satisfy the commutation relations 
\equ{
[M_{ab}, M_{cd}] = \delta_{ac}M_{db} - \delta_{bd}M_{ac} - 
\delta_{ad}M_{cb} + \delta_{bc}M_{ad}
\labl{so}.
}
We denote the $N^2$ generators of $U(N)$ by $U^i\,_j$
$(i ,j = 1,\ldots, N)$. The remaining $\smash{N(2N-1) - 
N^2 = N(N - 1)}$ generators form two anti-symmetric tensor 
representations of $U(N)$: $X^{ij}$ and $\bar{X}_{ij}$, each 
of dimension $N(N-1)/2$. The $U(N)$ generators satisfy the 
algebra 
\begin{equation} [U^i\,_j,U^k\,_l] = \delta^i\,_lU^k\,_j -
\delta^k\,_jU^i\,_l. \labl{un}
\end{equation}
\par

We decompose  the $SO(2N)$ algebra w.r.t.\ $U(N)$ by writing the 
$SO(2N)$ generators $M_{ab}$ using indices 
$i, j = 1, \ldots, N$ as 
\equ{
\barr{lcc}
M_{ij} & =& \frac{1}{2}(-X^{ij} - \bar{X}_{ij} - U^i\,_j 
 + U^j\,_i), 
\\[2mm]
M_{i\,j+N} &=& \frac{i}{2}(X^{ij} - \bar{X}_{ij} - U^i\,_j 
 - U^j\,_i),
\\[2mm]                  
M_{i+N\,j+N} &=& \frac{1}{2}(X^{ij} + \bar{X}_{ij}  - U^i\,_j + 
 U^j\,_i).
\earr
\labl{sy-anti}
}
Inversely we can express  $U^i\,_j, X^{ij}$ and $\bX_{ij}$ as 
$U^i\,_j = A^i\,_j + i S^i\,_ j$ with 
\equ{
A^i\,_j =  - \half \lh M_{ij} + M_{i+N\, j+N} \rh,
\qquad 
S^i\,_j = \half \lh M_{i\, j+N} + M_{j\, i+N} \rh
\labl{AandS}
}
and 
\(
X^{ij}= - iQ^{ij} - P^{ij} 
\) and 
\(
\bar{X}_{ij} = iQ^{ij} - P^{ij}
\)
with
\equ{
P^i\,_j = \frac{1}{2} \lh M_{i\,j} - M_{i+N\,j+N} \rh,
\qquad
Q^i\,_j = \frac{1}{2} \lh M_{i\,j + N} - M_{j\,i+N} \rh.
\labl{PandQ}
}
The U(1)-factor generator Y in $U(N)$ is defined as minus twice 
the trace of the $U(N)$ generators 
\equ{
Y = -2\sum^N_iU^i\,_i = -i2S^i\,_i = -2iM_{i\,i+N}
\labl{u1}
}                                                           
and the remaining SU(N) generators $T^i\,_j$ are define as the 
traceless part of $U^i\,_j$
\equ{
T^i\,_j= U^i\,_j + \frac{1}{2N}Y\delta^i\,_j.
} 
Using the $U(N)$ generators $U^i\,_j$ and the broken generators 
$X^{ij}$ and $\bX_{ij}$ the $SO(2N)$ algebra \eqref{so} takes the 
form 
\equ{
\mtrx{
\mtrx{
[U^i\,_j,U^k\,_l]= \delta^i\,_lU^k\,_j - \delta^k\,_jU^i\,_l,
&
\quad [{X}^{ij}, X^{kl}] = [\bar{X}_{ij}, \bar{X}_{kl}]= 0,
}
\\ \\ \
[\bar{X}_{ij},X^{kl} ] =  -\delta^k\,_iU^l\,_j - 
 \delta^l\,_jU^k\,_i + \delta^l\,_iU^k\,_j + \delta^k\,_jU^l\,_i,
\\ \\ \   
[U^i\,_j,\bar{X}_{kl}] =  \delta^i\,_k\bar{X}_{jl} - 
\delta^i\,_l\bar{X}_{jk} ,
\\ \\ \
[U^i\,_j,X^{kl}  ] = \delta^l\,_jX^{ik} - \delta^k\,_jX^{il} .
}
\labl{alg}
}     
The closure of the algebra can be checked explicitly by computing 
the Jacobi identities. The $SO(2N)$ generators in this basis carry 
the following $U(1)$-charges:
\equ{
U(1)\text{-charges of }( Y, T^i\,_j, X^{ij}, \bX_{ij} ) = 
 ( 0,  0, \,4,\, -4 ).
}
Here we have chosen the $U(1)$-charges such that they match the 
convention of Slansky \cite{Slansky:1981yr}.
%
%
\subsection{The vector representation of $\bf{SO(2N)}$ }
\label{vector_representation}

In the vector representation of $SO(2N)$, the generators $M_{ab}$ 
take the form: $(M_{ab})_{cd} = \gd_{ac}\gd_{bd} - 
\gd_{bc}\gd_{ad}$, therefore an element of the $SO(2N)$-algebra 
reads
\equ{
\gTh = (-a^{ij}A_{ij} - s^{ij}S_{ij}) + (q^{ij}Q_{ij} - 
p^{ij}P_{ij}) \equiv \left(
\mtrx{
a&-s\\
s&~~a
}
\right) + \left(
\mtrx{
-p&q\\
~~q&p
}
\right),
\labl{aspq}
}
where $a$, $p$, $q$ are $N\times N$ real anti-symmetric matrices 
and $s$ is a real symmetric $N\times N$ matrix; these matrices
define the parameters of the $SO(2N)$-algebra elements. Here we 
have used the definitions of the algebra elements $A, S, P$ and 
$Q$ given in eqs.\ \eqref{AandS} and \eqref{PandQ}. The 
U(1)-factor generator Y \eqref{u1} in the vector representation 
takes the form 
\equ{
Y = -2i\left(
\mtrx{
~~0&\Id\\
-\Id&0 
}
\right)
\labl{ucharge}.
} 
Notice that the $U(1)$ generator $Y$ is off-diagonal. 
However it is more convenient to use  a basis in which $Y$ is 
diagonal. Using a unitary transformation we can diagonalize $Y$:
\equ{
Y_D \equiv V Y V^\dag = 2 \pmtrx{ \Id & ~~0 \\ 0 & -\Id }
\quad \text{with} \quad
V = \frac{1}{\sqrt{2}}\pmtrx{ ~~~\Id & -i \Id \\ -i \Id & ~~~\Id}.
}
We use the subscript notation $D$ on any $2N\times 2N$-matrix 
$A$ to indicate that $A$ is evaluated in the basis where $Y$ is 
diagonal. The effect of this similarity transformation on an 
element $\gTh$ of the $SO(2N)$ Lie algebra \eqref{aspq} is given 
by
\equ{
\gTh_D = VMV^\dagger = 
\left(
\mtrx{
a - is & q - ip \\
q + ip & a + is
}
\right) = 
\left(
\mtrx{
u& -x^\dag \\
x&-u^T
}
\right),
\labl{gThD}
}
where 
$u = - u^\dag =  a - is$, $u^T = a + is$, $x = q + ip$ and 
$x^\dag = -q + ip$. This is coincides with the $SO(2N)/U(N)$ 
entry in table \ref{EmbeddingSU}. Notice that in the basis where 
$Y$ is diagonal, the defining property $g\inv = g^T$ of $SO(2N)$ 
becomes
\equ{
g_D\inv = \fK g_D^T \fK
\quad \text{with} \quad 
\fK \equiv \pmtrx{ 0 & \Id \\ \Id & 0 }.
\labl{modSOgroup}
}
Writing $g_D$ in terms of the submatrices $\ga, \gb, \gg$ and 
$\gd$ as introduced in eq.\ \eqref{Mtrxg} this group property can 
be stated as 
\equ{
\pmtrx{ \siga & \sigb \\ \sigg & \sigd} = 
\pmtrx{ \gd^T & \gb^T \\ \gg^T & \ga^T}, 
\labl{InvisTransp}
}
using the notation \eqref{Mtrxinfg} for the inverse of $g_D$.
From now on we will only work in the basis where the $U(1)$-charge 
$Y$ is diagonal, dropping the subscripts $D$. 

%
%
\subsection{\Kh\ and Killing Potentials}
\label{KaehlerKilling}

We now construct the \Kh\ potential for the coset spaces 
$SO(2N)/U(N)$ using the BKMU-method \cite{BKMU}. We apply 
their method to the $2N$ dimensional  vector representation of 
$SO(2N)$. The BKMU-projection $\get_+$ projects \eqref{projpm} 
on the part of this vector representation with positive $Y$-charge, 
which is an $N$ dimensional vector representation of $SU(N)$. The 
coset spaces $SO(2N)/U(N)$ and $SO^\Cplx(2N)/\hat{U}(N)$, with 
$SO^\Cplx(2N)$, the complexification of $SO(2N)$, are isomorphic 
because $\hat U(N)$ is defined as the group generated by all 
generators of $U(N)$ together with the broken generators $X^{ij}$ 
over the complex numbers. The representative 
\(
\gx(z) \in SO^\Cplx(2N)/\hat{U}(N) \cong SO(2N)/U(N)
\)
of the equivalence class $\gx(z)  \hat U(N)$ is given in terms of 
the $\half N(N-1)$ coordinates $z^{ij}$ of $SO(2N)/U(N)$ by 
\equ{
\xi(z) = \exp Z = 
\left(
\mtrx{
\Id&0\\
z&\Id
}
\right), 
\qquad Z = -\frac{i}{2}z^{ij}\bar{X}_{ij}.
\labl{BKMU}
}
On the r.h.s.\ of the equation for $\gx(z)$ we used the vector 
representation in the diagonal $U(1)$-charge Y basis, where $Z$ 
is nilpotent $Z^2 = 0$.  The normalization factor $-\frac{i}{2}$ 
in the definition of $Z$ is chosen such that we get the simple 
matrix expression for $\gx(z)$ expressed in terms of $z$ which 
coincides with \eqref{DefXiSL}. Notice the distinction between 
$z$ and $Z$: $Z$ is the linear combination of negatively charged 
broken generators $\bar{X}_{ij}$ contracted with the complex 
coordinates $z^{ij}$ of the coset space. Therefore $Z$ is 
represented by a $2N\times 2N$ matrix, while $z$ is an $N\times N$ 
matrix. Using the projection operator $\get_-$ defined in eq.\ 
\eqref{projpm} and $\gx(z)$ the \Kh\ potential is given by 
\eqref{CanonicalKh} and \eqref{tChiI}
\equ{
K(z,\bar{z}) = \ln\det {}_{\eta_-}[\xi(-z)\xi^\dagger(-\bar{z})] 
=  \ln\det \gch\inv, 
\qquad \gch\inv = \Id + z \bz.
\labl{kahler}
}
Here the $\det{}_{\eta_-}$ denotes that the determinant is 
defined on the subspace on which the projection $\get_-$ acts 
as the identity. Notice that the submetric $\tgch$ defined in 
\eqref{ChiI} is the transposed $\tgch = \gch^T$ of $\gch$ because 
of the anti-symmetry of $z$. 

We next determine the non-linear transformations of the 
anti-symmetric coordinates $z^{ij}$ under the finite $g \in 
SO(2N)$ transformation. From eq.\ \eqref{transz} we know  
directly that 
\equ{
^gz = (\gg + \gd z)(\ga +\gb z)^{-1}.
\labl{global}
}
The submetric $\gch$ transforms under these finite 
$SO(2N)$-transformations as 
\equ{
\qquad
\gch(^g z, ^g \bz) = (\hat h_-^\dag)\inv \gch(z, \bz) 
(\hat h_-)\inv, \qquad
\hat h_-(z;g) = (\sigd - z \sigb)\inv,
\label{SubMetric}
}
using eq.\ \eqref{hath}. Notice that according to eq.\ 
\eqref{InvisTransp} $\hat h_+(z;g) = \hat h^T_-(z;g)$ is the 
transposed of $\hat h_-$ , therefore we only use $\hat h_-$ in 
the following. The \Kh\ potential \eqref{kahler} transforms as 
follows 
\equ{
K(^g z, ^g \bar{z}) = K(z,\bar{z}) + F(z; g) + 
\bar{F}(\bar{z}; g),
}
where the holomorphic function $F(z; g)$ is given by
\equ{
F(z; g) = \ln\det\hat{h}_-(z;g) 
\labl{holomorphic}
}
The complex Hermitean metric of the coset is obtained from the  
\Kh\ potential \eqref{kahler} in the standard way as the second 
mixed derivative
\equ{
G_\gs (d z, d \bz) = \tr \left(
d z\,  \gch^T \,  d \bz \, \gch
\right)
=
\tr \left( d z \, \bigl( \Id + \bz z \bigr)\inv
 \, d \bz\, ( \Id + z \bz)\inv \right).
}
We next discuss the Killing potentials $M_\gs$ for the Goldstone 
scalar fields $z$ and $\bz$. The Killing potential $M_\gs$, 
defined by eq.\ \eqref{Killing_Pot}, can be written for the coset 
$SO(2N)/U(N)$ as
\equ{
M_\gs(u, x, \bx) = \Tr(\gTh \tM_\gs) = \tr (uM_{\gs\,u}+ xM_{\gs\, 
 x^\dag} + x^\dag M_{\gs\, x}),
}
where the trace $\Tr$ is over $2N\times 2N$ matrices, while the 
trace $\tr$ is over $N\times N$ matrices. We have used a 
notation similar to eq.\ \eqref{gThD} 
\equ{
\gTh = \pmtrx{ u & -x^\dag \\ x & - u^T }
\quad\text{and}\quad
\tM_\gs = 
\pmtrx{
~~\tM_{\gs\, u} & ~~\tM_{\gs\, x^\dag}
\\
-\tM_{\gs\, x} &-\tM_{\gs\, u^T}
},
\labl{theta}
}
so that
\(
M_{\gs\, x} = \tM_{\gs\, x}, ~ 
M_{\gs\, x^\dag} = \tM_{\gs\, x^\dag}
\)
and 
\(
\smash{M_{\gs\, u} = \tM_{\gs\, u} + \bigl( \tM_{\gs\, u^T}
 \bigr)^T.}
\) 
We now determine the Killing potentials explicitly. We will 
introduce some notation that might seem somewhat cumbersome at 
this stage, but which will be convenient when we discuss the 
Killing potentials due to additional matter coupling. Define the 
matrices $R$ and $R_T$ by
\equ{
R(z; \gTh)  = x - u^Tz - zu + zx^\dag z, 
\qquad 
R_T(z; \gTh) = -u^T + zx^\dag.
\label{RandRT}
}
Notice that $\gd z = R(z; \gTh)$ is a compact notation for the 
Killing vectors of the coset space, and $\tr R_T = F(z)$, the 
holomorphic K\"{a}hler transformation. Computing the Killing 
potentials $M_\gs$ in the standard way (\ref{Killing_Pot}) 
gives
\equ{
-i M_\gs(z, \bz; \gTh) = - \tr\, \gD(z, \bz; \gTh),
\label{MgD}
}
where we have defined the matrix $\gD$ in analogy to the Killing 
potentials associated with the Grassmannian cosets \cite{GNvH2} 
by  
\equ{
\gD(z, \bz; \gTh) \equiv R_T - R\bz\gch  =  
(zu \bz - u^T  - x \bz + zx^\dag) \gch.
\label{gD}
}
The matrix $\gD$ can also be written in terms of the 
BKMU-variable $\gx(z)$ and $\gTh$ and the projector 
\(
\tget_-^T = \pmtrx{ 0 & \Id }
\)
as 
\equ{
\gD(z, \bz; \gTh) = \tget_-^T \lh \gx(z) \rh\inv 
\gTh \lh \gx^\dag(\bz) \rh\inv\tget_-\, \gch.
\label{gDingTh}
}
Using that $\lh \gx(z) \rh\inv = \gx(-z)$,  
the Killing potential matrix $\tM_\gs$ is given by
\equ{
-i \tM_\gs = - 
\gx^\dag(-\bz) \tget_-\, \gch\, 
\tget_-^T \gx(-z)  
= -  
\pmtrx{
~\bz\gch z & - \bz\gch
\\
- \gch z & ~~\gch
}.
}
From this we can read off the Killing potentials $M_{\gs\, x}, 
M_{\gs\, x^\dag}$ and $M_{\gs\, u}$ to find
\equ{
-i M_{\gs\, x} = -\gch z, 
\quad
-i M_{\gs\, x^\dag} =  \bz \gch, 
\quad 
-i M_{\gs\, u} = -2 \bz \gch z + \Id.
\labl{SigmaKilling}
}

%
%
\subsection{Matter coupling }
\label{MatterCoupling}

In this section we discuss different types of matter couplings to 
the supersymmetric $SO(2N)/U(N)$ $\gs$-model. As we only need the 
decomposition of the chiral spinor representation of $SO(2N)$ in 
completely anti-symmetric $SU(N)$ tensors in our construction of 
anomaly-free models later, we focus here primarily on these 
representations. We first introduce a matter representation $x$ 
which transforms in the same way as a differential. Under a finite 
transformation \eqref{global} the real superfield $x$ transforms 
as
\equ{
^g x = \hat{h}_-(z; g)x\hat{h}_-^T(z; g),
}
using that $\hat h_+ = \hat h_-^T$. 
An invariant \Kh\ potential for $x$ is given by
\equ{
K(x,\bx;z,\bz) = \tr \left( x\gch^T \bx\gch \right).
}   
Below we discuss non-linear $SO(2N)$ realizations on the 
irreducible completely anti-symmetric $SU(N)$-tensor 
representations with $p$ indices and arbitrary rescaling charge 
$q$. We denote these tensors by $T_{(p; q)}^{i_1\dots i_p}$, 
or without indices by $T_{(p;q)}$, when no confusion is
possible. We interpret them as matter multiplets and construct 
their invariant \Kh\  potentials. To define their transformation 
properties we first consider a vector $T^i = T^i_{(1; 0)}$ 
without a rescaling charge. It transforms as 
\equ{
^g T = \hat{h}_-(z; g)T,
}
under finite non-linear $SO(2N)$ transformations \eqref{global}. 
An invariant \Kh\ potential for the vector $T = T_{(1;0)}$ is 
given by
\equ{
K_{(1;0)} = \bT \gch T = \bT_i\,\gch^i\,_j T^j,
\labl{kahlerT}
}
with the metric $\gch$ defined in eq.\ \eqref{kahler}. 

It is also possible to couple a singlet chiral multiplet $S$ to 
the coset, which can be interpreted as a section of the minimal 
line bundle. It transforms as \eqref{LineUSpSO}  
\equ{
^gS = e^{\frac{1}{2}F(z)}S = (\det\hat{h}_-) {}^{\frac{1}{2}}S, 
\labl{transfors}
}
so that its \Kh\ potential 
\equ{
K_{(0; 1)} =  S\bS e^{-\frac{1}{2} K_\gs}
\labl{kahlerS}
}
is invariant. With this singlet $S$, we can rescale any given 
chiral multiplet, for example $T^i_{(1; q)} \equiv 
S^qT^i_{(1; 0)}$ transforms as
\equ{
^gT_{(1; q)} =\,  ^g(S^qT_{(1; 0)}) = e^{\frac{q}{2}F(z)}\hat{h}_-
T_{(1; q)} = (\det\hat{h}_-) {}^{\frac{q}{2}}\hat{h}_-T_{(1; q)}.
}
Since $S$ is a section of the minimal line bundle over the 
coset $SO(2N)/U(N)$ the rescaling charge $q$ is integer. The 
generalization of \Kh\ potential \eqref{kahlerT} is given by
\equ{
K_{(1; q)}= \bT_{(1; q)}\gch_{(1; q)}T_{(1; q)},
}
with the modified metric
\equ{
\gch_{(1; q)} = e^{-\frac{q}{2}K_{\gs}}\gch = 
(\det \gch) ^{\frac q2} \gch.
}
Now we construct completely anti-symmetric tensor representations 
of higher rank. By taking the completely anti-symmetric  tensor 
products of a set of $SU(N)$ vectors $\{ T_1^{i_1}, \ldots, 
T_p^{i_p} \}$ we obtain an $SU(N)$ tensor of rank $p$ with 
rescaling charge $q$ 
\equ{
T_{(p; q)}^A = T^{i_1\ldots i_p}_{(p;q)} \equiv
\frac 1{p!} S^q T^{[i_1}_1* \ldots * T^{i_p]}_{p}.
}
Here we have introduced the multi-index notation $A = (i_1 \ldots 
i_p)$ and $[\ldots]$ denotes the complete anti-symmetrization 
of the indices inside the brackets. In analogy to the 
transformations of $T_{(1;0)}$ and $S$ we obtain 
\equ{
^gT^{i_1\dots i_p}_{(p; q)} = 
(\det\hat{h}_-) {}^{\frac{q}{2}}(\hat{h}_-)
^{i_1}\,_{j_1}\dots (\hat{h}_-)^{i_p}\,_{j_p}
T^{j_1\dots j_p}_{(p; q)}.
\labl{TransTensor}
}
The \Kh\ potential for this tensor $T_{(p; q)}$ is the direct 
generalization \cite{GNvH1} of the \Kh\ potentials for the vector 
\eqref{kahlerT} and singlet \eqref{kahlerS}   
\equ{
K_{(p; q)} = \bar{T}_{(p; q)B} \,  G^B_{(p; q)\,A} \, T^A_{(p; q)} 
= \frac{1}{p!} \bar{T}_{(p; q)j_1\dots j_p}\,
e^{-\frac{q}{2}K_{\gs}}\, \gch^{j_1}_{~i_1}\dots 
\gch^{j_p}_{~ i_p}\, T^{i_1\dots i_p}_{(p; q)},
\labl{KahlerTensor}
}
with the generalized metric 
\equ{
G^B_{(p; q)\,A} =\frac{1}{p!}(\det \gch)^{\frac{q}{2}}
\gch^{j_1}_{~i_1}\dots \gch^{j_p}_{~ i_p}.
\labl{rankptensor}
}
The $SU(N)$ Levi-Civita tensor $\ge_{i_1\dots i_N}$ is invariant 
under $SU(N)$ transformations. We can use it to defined an $SU(N)$ 
dual tensor $T_{(\ovr{N-p}; q)\, i_{p+1}\dots i_{N}}$ with $N-p$ 
indices and rescaling charge $q$ by 
\equ{
T_{(\overline{N-p}; q)\, i_{p+1}\dots i_{N}} \equiv 
\frac{1}{p!}T_{(p; q)}^{i_p\dots i_1} \ge_{i_1\dots i_N}, 
}
which transforms under the finite transformation \eqref{global} as
\equ{
^g T_{(\overline{p};q) i_1\ldots i_{p}} = 
T_{(\overline{p};q) j_1 \ldots j_{p}}
(\hat{h}_-\inv)^{j_1}\,_{i_1} \ldots (\hat{h}_-\inv)^{j_p}\,_{i_p}
(\det \hat{h}_-)^{1+\frac q2}.
\labl{TransDualTensor}
} 
Note, that as $\sqrt{G}$ is not holomorphic, we have prefered to 
absorb it in a redefinition of the metric, rather than in the 
definition of the dual. The power $\smash{1 + \frac q2}$ of the 
$\det \hat h_-$ instead of $\smash{\frac q2}$ arises because we 
have changed from $\hat h_-$ to its inverse at the expense of 
an additional factor of the determinant of $\hat h_-$. In our 
conventions tensors have superscript indices while dual tensors 
have subscript indices. Cleary, working with anti-symmetric 
tensors or dual tensors is equivalent. The invariant \Kh\ 
potential for a dual tensor is given by 
\equ{
K_{(\bp; q)} =
T_{(\overline{p}; q)\, A}\,
G^A_{(\overline{p}; q)\,B} \,
\bar{T}_{(\overline{p}; q)}^B 
\labl{KahlerDualTensor}
}
where the metric is given by 
\equ{
G^A_{(\overline{p}; q)\,B}  = \frac{1}{p!}(\det \gch)^{1 + 
\frac{q}{2}} (\gch\inv)^{i_1}_{~j_1}\dots (\gch\inv)^{i_p}_{~j_p}.
} 
In addition we can construct a matter reprensentation $A$ that 
transforms in the adjoint of $SU(N)$. The index structure of this 
matrix is $A^i_{\; j}$ and in addition it is traceless 
$\tr A = A^i_{\; i} = 0$. It transformation properties 
under the full non-linear $SO(2N)$ symmetries takes the form
\equ{
^g A = \hat h_-(z; g) A \hat h_-\inv(z;g).
\label{transA}
}
This transformation rule can be obtained by defining $A$ as 
the tensor product of a vector $T_{(1;0)}$ and a dual vector 
$T_{(\bar 1; -2)}$ with rescaling charge $-2$ 
\equ{
A = T_{(1;0)} \otimes T_{(\bar 1; -2)}.
}
It is easy to see that this gives the right generalization 
of the $SU(5)$ adjoint by restricting $SO(2N)$ to an $U(5)$ 
transformation:
\equ{
g = \pmtrx{
\ga & 0 \\ 0 & (\ga^T)\inv
}
\Lra
{}^g A = (\ga^T)\inv A \ga^T. 
}
Clearly, $A$ does not transform under the $U(1)$ factor of 
$U(5)$. Notice that the condition that $A$ be traceless, 
is respected by the transformation rule \eqref{transA}. 
The simplest invariant \Kh\ potential for this matter 
field $A$ is
\equ{
K_A = \tr \left( \gch A \gch\inv \bA \right).
}
We next turn to a discussion of the contributions $M_{(p; q)}$ 
and $M_{(\bp; q)}$ to the Killing potentials for a tensor 
$T_{(p; q)}$ and a dual-tensor $\bT_{(\bp; q)}$ of rank 
$p$ with a rescaling charge $q$, respectively. As the \Kh\ 
potentials $K_{(p; q)}$ and $K_{(\bp; q)}$ are invariant,
their contributions to the Killing potentials are obtained 
from   
\equ{
-iM_{(p; q)} = K_{(p; q),\,\ga} \cR^\ga, 
\qquad
-iM_{(\bp; q)} = K_{(\bp; q),\,\ga} \cR^\ga,
\labl{Matterk}
} 
where $\gd_i Z^\ga = \cR^\ga_i$ denote the Killing vectors 
(cf.\ eq. (\ref{RandRT}))
\equa{
\gd z &= R,
\non \\[2mm]
\gd T^{i_1\dots i_p}_{(p; q)} & = {\dsp \sum_{r=1}^p} 
 (R_T)^{i_r}{}_j T^{i_1\ldots  j\ldots i_p}_{(p; q)} 
 + \frac{q}{2} \,\tr (R_T) \, T^{i_1\dots i_p}_{(p; q)},
\labl{vKilling}\\[2mm]
\gd T_{(\bp; q)i_1\dots i_p} & =
{\dsp \sum_{r=1}^p} T_{(\bp; q)i_1\ldots  j\ldots i_p} 
 (-R_T)^j{}_{j_r} + \Bigl( 1 + \frac q{2}\Bigr) \tr (R_T)\, 
 T_{(\bp; q)\, i_1\dots i_p}.
\non 
}
They follow from expanding the finite transformations 
\eqref{global}, \eqref{TransTensor} and \eqref{TransDualTensor}
to first order in the infinitesimal parameters $u, x, x^\dag$. 

The Killing potential for a rank $p$ tensor with rescaling charge 
$q$ is given by
\equ{
-iM_{(p; q)} = \bT_{(p; q)B}G^B_{(p; q)C}\gD^C_{(p; q)A}
 T^A_{(p;q)},
\labl{KillingTensor}
}
where, using the notation (\ref{gD}), 
\equ{
\gD^C_{(p; q)A} = 
\sum_{r=1}^p\gd^{k_1}{}_{i_1}\ldots\gD^{k_r}{}_{i_r}\ldots
\gd^{k_p}{}_{i_p} + \frac{q}{2}\tr\gD\, 
\gd^{k_1}{}_{i_1}\ldots \gd^{k_p}{}_{i_p}.
}
To obtain this result we have made the following steps. We first 
obtained the Killing potential for a rank 1 tensor (a vector) 
with rescaling charge zero. This result can easily be generalized 
to a rank $p$ tensor with rescaling charge zero. Next we 
construct the Killing potential for a rank 0 tensor (a singlet) 
with an arbitrary rescaling charge. Finally we put all results 
together to obtain eq.\ \eqref{KillingTensor}. 

\nit
We can proceed similarly to obtain the Killing potential 
$M_{(\bp; q)}$ for a rank $p$ dual tensor with a rescaling charge 
$q$. As the dualization has introduced a determinant $\det \hat 
h_-$ in the finite transformation \eqref{TransDualTensor}, it is 
more convenient to first consider a rank $p$ dual tensor with 
rescaling charge $-2$, which precisely cancels the determinant. 
To obtain the final result for a rank $p$ dual-tensor with a 
rescaling charge $q$, we have to rescale the rank again, which 
introduces a factor $1 + \smash{\frac q2}$. Finally, the Killing 
potential reads 
\equ{
-iM_{(\bp; q)} = 
T_{(\bp; q)B}\gD^B_{(\bp; q)C} G^C_{(\bp; q)A} \bT^A_{(\bp; q)},
\label{KillingDualTensor}
}
with $\gD^C_{(\bp; q)A}$ defined as
\equ{
\gD^B_{(\bp; q)C} = 
\sum_{r=1}^p\gd^{j_1}{}_{k_1}\ldots(-\gD)^{j_r}{}_{k_r}\ldots
\gd^{j_p}{}_{k_p} + \Bigl(1 + \frac{q}{2}\Bigr) \tr\, \gD\, 
\gd^{j_1}{}_{k_1}\ldots\gd^{j_p}{}_{k_p}. 
} 
The infinitesimal form of the transformation of the adjoint matter 
field $A$ is given by
\equ{
\gd A = R_T A - A R_T = [ R_T, A]
}
and the resulting Killing potential can be written as 
\equ{
- i M_A = \tr \left( 
\gch \gD A \gch\inv \bA - \gch A \gD \gch\inv \bA
\right) = \tr \left( \gch [ \gD,  A]  \gch\inv \bA \right).
}

%
%
\subsection{Consistent $\boldsymbol{SO(2N)/U(N)}$ spinor models
  \label{s5}}
\label{ConsistentSpinorModels}

In this subsection we construct an anomaly-free model based 
on the spinor representation of $SO(2N)$ that contains the 
coordinates of the coset $SO(2N)/U(N)$. Only for a limited 
number of choices for $N$ such a model satisfies the line 
bundle constraint. 

A supersymmetric model built on the $SO(2N)/U(N)$ coset 
space is not free of anomalies by itself, as all the $\half 
N(N-1)$ anti-symmetric coordinates $z^{ij}$ and therefore 
also their chiral fermionic partners carry the same charge 4 
in the standard normalization. To construct a consistent 
supersymmetric model around this coset one can try to embed 
the coordinates in an anomaly-free representation. All 
representations of $SO(2N)$ are anomaly-free, unless $SO(2N)$ 
is isomorphic to a non-anomaly-free unitary group. This 
happens for $SO(2) \cong U(1)$ and $SO(6) \cong SU(4)$, 
hence we disregard the cases $N = 1, 3$ below. In appendix 
B we derive this result by calculating the possible 
$U(1)$ anomalies of the chiral spinor representation. An 
$SO(2N)$ representation that branches to an anti-symmetric 
$2$-tensor of $SU(N)$ is the chiral spinor representation 
of $SO(2N)$. The other $U(N)$ representations that arise 
from the spinor representation transform under the full 
$SO(2N)$ symmetries via non-linear transformations. For 
global consistency this means that these matter 
representations are sections of bundles. If one of these 
sections is a line bundle we run into the cocycle 
condition, which greatly restricts the freedom of charge 
assignments. In section \ref{SLNC} we have determined the 
section of the minimal line bundle over $SO(2N)/U(N)$. As 
the dimension $2N$ is even, the irreducible representations 
carry definite chirality; we show that it is sufficient to 
consider only the positive chiral spinor representation 
for our purpose of extending the coset to the spinor 
representation. After that we turn to the main result of 
this subsection: the cocycle condition only allows for a 
very restricted class of consistent $SO(2N)/U(N)$ spinor 
models: $N = 2, 5, 6, 8$.

As argued in appendix A, to construct a consistent model on  
$SO(2N)/U(N)$ using irreducible spinor representations, we 
need to identify the anti-symmetric coordinates $z^{ij}$ of 
the coset space with an anti-symmetric $2$-tensor of the 
branching of the spinor. We have the following two states 
$\gps_2^{~ij}$ or $\gps_{\bar 2\,ij}$ as  possible 
candidates. According to eq.\ (\ref{Ycharge}), appendix A, 
the charge of $\gps_2^{~ij}$ is $N-4$; it has positive 
chirality. The charge of $\gps_{\bar 2\, ij}$ is opposite 
and its chirality is $(-)^N$. Notice that for $N = 4$ we 
can never construct a consistent model using the spinor 
representations as the charges of $\gps_2^{~ij}$ and 
$\gps_{\bar 2\, ij}$ are zero, while the charge of the 
coordinate $z^{ij}$ is non-zero. For $N$ is even both 
$\gps_2^{~ij}$ and $\gps_{\bar 2\,ij}$ have the same 
chirality, hence they are in the same irreducible 
representation. The duality operation \eqref{DualOp}, 
appendix A, maps the positive chirality states into 
themselves. Therefore, for even $N$ it is sufficient 
to consider only the state $\gps_2^{~ij}$ as the candidate 
for the coordinates $z^{ij}$ of the coset. For odd $N$ the 
only odd length state that can be associated with the 
coordinates $z^{ij}$ has length $N-2$, but it is dual to 
the state with length $2$. Therefore, for all $N$ it is 
sufficient to consider only the positive chirality spinor 
representation and only the state $\gps_2^{~ij}$ as 
candidate for the coordinates $z^{ij}$ of $SO(2N)/U(N)$. 

We next discuss the restriction that the consistency of 
the line bundle poses on the construction of anomaly-free 
extensions of cosets $SO(2N)/U(N)$ using the positive 
chirality spinors of $SO(2N)$. We remarked before that 
the case $N = 4$ does not work as the state $\gps_2^{~ ij}$ 
does not carry $Y$ charge. Therefore we consider the cases 
$N = 2$ and $N \geq 5$ from now on. It was shown in section 
\ref{SLNC} that the minimal charge of the line bundle over 
the coset space $SO(2N)/U(N)$ is equal to $N$ when the 
charge of the coordinates is taken to be $4$. This is the 
normalization employed in our detailed discussion of the 
$SO(2N)$ algebra in sect.\ \ref{SOinU}. As all states of a 
positive chirality spinor have an even number of indices, 
the tensor structure of these states can be obtained from 
completely anti-symmetric tensor products of the tangent 
vectors of the coset $SO(2N)/U(N)$ tensored with an integral 
power of the minimal line bundle. In particular the state 
$\gps_{(2p;q)}$ with length $2p$ and rescaled with the 
$q(p; N)$th  power of the minimal line bundle has a 
charge $4p + Nq(p;N)$. For each $p$ this charge should be 
proportional to the charge $N - 4 p$ of the anti-symmetric 
tensor with $2p$ indices within the positive  chirality 
spinor representation. Therefore we obtain the relation 
\(
\gl (N - 4 p) =  4p + N q(p; N) 
\)
where $\gl \in \Real$ is a constant to be determined. Since the 
anti-symmetric tensor with 2 indices ($p = 1$) is identified 
with the coordinates $z^{ij}$ of the coset, it does not have a 
rescaling charge, hence we find that $ \gl = \frac {4}{N - 4}$. 
Solving for $q(p;N)$ gives
\equ{
q(p;N)\, =\, q(0;N)\, \lh 1 - p \rh\, =\, \frac 4{N-4}\, (1 - p ).
\label{ReChargeSpin}
}
For consistency of the line bundle we need that $q(p;N)$ is an 
integer for all $0 \leq p \leq [N/2]$. Notice that $q(p;N)$ is 
integer whenever $q(0;N)$ is integer. $q(0; N)$ is only an 
integer if $N-4$ is a divider of $4$, which implies that $N = 0, 
2, 3, 5, 6, 8$. Of course $N = 0$ is impossible, and though the 
case $N = 3$ satisfies the line bundle quantization condition, 
it does not lead to an anomaly free model. Therefore the 
possible choices are:
\[
\renewcommand{\arraystretch}{1.5}
\begin{array}{|l | c | c | c | c |}
\hline
N                   &  2 &  5 &  6 &  8 
\\ \hline
q(0; N) = \frac {4}{N-4} & -2 & 4 & 2 &  1 
\\ \hline
\end{array}
\]
The case of $N = 2$ is trivial in the sense that the coset 
is isomorphic to the simplest coset $SU(2)/U(1)$ (i.e., the 
2-sphere) because
\(
SO(4) \cong SU(2) \times SU(2).
\)
Notice that except for the last case $N = 8$ we only use 
squares of the minimal line bundle.  

We finish  this section by giving the \Kh\ potentials for the 
anomaly-free $SO(2N)/U(N)$ models based on the positive chiral 
spinor representation. The matter content is fixed by the 
discussion above: we need for each $0 \leq p \leq [N/2]$ a rank 
$2p$ completely anti-symmetric $SU(N)$ tensor with rescaling 
charge $q(p; N)$ given in eq.\ \eqref{ReChargeSpin}, except for 
$p = 1$; this case corresponds to an anti-symmetric tensor with 
two indices, for which we take the coordinates of the 
$SO(2N)/U(N)$ coset itself. Using the \Kh\ potentials for 
the coset \eqref{kahler} and for anti-symmetric tensor 
representations with an arbitrary rescaling charge 
\eqref{KahlerTensor}, we can express the \Kh\ potential for the 
complete system by
\equ{
\cK = \half K_\gs + 
\sum_{\scp p = 0,\  p \neq 1}^{\scp [N/2]} K_{(\,2p;~ q(p; N) \,)}.
\labl{FullKahlerSpin}
}
Here we have included a factor $\half$ so as to get the standard 
normalization of the kinetic terms of the Goldstone boson fields. 
In section \ref{PhenofSO(10)} we discuss the consistent 
$SO(10)/U(5)$-spinor model in detail. There we give the explicit 
expression for the \Kh\ potential, using dual tensors to reduce 
the number of indices. 

\newcommand{\pl}{\partial}  

\section{Analysis of the $\bf{SO(10)/U(5)}$-spinor 
model \label{s6}} \label{PhenofSO(10)}

In this section we apply the constructions presented 
so far to obtain anomaly-free ${SO(10)/U(5)}$-spinor 
models in the context of global supersymmetry. We study 
in particular the role of the potentials in determining 
the realization of internal symmetries and supersymmetry 
in the original $\gs$-model, as well as in various 
gauged versions. We find some rather surprising results 
concerning the gauged versions of the model, implying
that ---in spite of selecting anomaly-free combinations 
of representations--- it is not possible to gauge just 
any arbitrary global symmetry. In particular, we find 
that gauging the full global $SO(10)$ is not possible, 
whilst the consistency of gauging all or part of the 
linear $SU(5) \times U(1)$ symmetry depends crucially 
on the vacuum expectation values and choice of parameters 
in the model. We present strong arguments that the natural 
value of the Fayet-Iliopoulos parameter $\xi$ in the models 
with a linear gauge group containing $U(1)$ is determined 
by the scale $f$ of the $\gs$-model, by a relation of the 
type $|\xi| f^2 \sim \cO(1)$. It is not difficult to see 
that some of these results are valid beyond the particular 
model chosen. A more general and extensive discussion will 
be given elsewhere \ct{prep}. 

The choice for the model on $SO(10)/SU(5) \times U(1)$ 
is motivated by its fermionic field content, corresponding 
to one complete family of quarks and leptons, including a 
right-handed neutrino. This can be seen by looking at the 
$SU(5)$ representations of the chiral multiplets that the 
model contains: the coordinate multiplets $\gF^{ij}$ form 
the $\undr{10}$ of $SU(5)$. The completely anti-symmetric 
tensor with 4 indices is equivalent to the $\undr{\bar{5}}$ 
with a relative $U(1)$ charge $-3$; we denote it by $\Psi_i$. 
And finally we have a singlet $\Psi$ of $SU(5)$, with $U(1)$ 
charge +5. 

We denote the full set of chiral superfields by $\gS^\ga 
= (\gF^{ij}, \Psi_i, \Psi)$, their physical components 
collectively by $(Z^\ga, \gps_L^\ga)$. The scalar components 
of the various $SU(5)$ representations are denoted by 
$Z^\ga = (z^{ij}, k_i, h)$. In the absence of any local 
gauge couplings, the kinetic part of the lagrangian for 
the model is given in terms of real composite superfields 
$\cK(\bar{\gS}, \gS)$ by the supersymmetric expression 
(\ref{1.7}), which is equivalent to 
\equ{
\mtrx{
\cL_\cK =  \cK(\bar{\gS}, \gS)|_D = - G_{\underline{\ga}\ga}
(\partial^\gm \bar{Z}^{\underline{\ga}}\partial_\gm Z^\ga  + 
\bar{\psi}^{\underline{\ga}}_L \stackrel{\leftrightarrow}{\sDer} 
\psi^\ga_L - \hat{H}^{\underline{\ga}}\hat{H}^\ga) \\ 
\\ 
\ +\frac{1}{2}\, R_{\underline{\ga}\ga\underline{\gb}\gb}
 (\bar{\psi}^\ga_R \psi^\gb_L) (\bar{\psi}^{\underline{\ga}}_L 
 \psi^{\underline{\gb}}_R). 
\label{so10l}
}
} 
In the present case, the \Kh\ potential from which the 
metric $G_{\underline{\ga}\ga}$ is derived, is given by
(\ref{FullKahlerSpin})
\equa{
\cK(\bar{Z}, Z) & = \half K_\gs + K_{(0; 4)} + K_{(\bar 1; -4)} 
\label{kp} 
\\[2mm]
& = \frac 1{2f^2} \ln\det \gch^{-1}+    
(\det \gch )^2|h|^2 + (\det \gch)^{-1}\, k\gch^{-1}\bk.
\non 
}
with the submetric 
\(
\gch\inv = \Id + f^2 z \bz
\)
and 
\(
\smash{e^{f^2 K_\gs} = (\det \gch)\inv}.
\)
This is the explicit form of eq.\ \eqref{FullKahlerSpin} in 
the $SO(10)/U(5)$ case, after rescaling the Goldstone fields 
$z$ by the mass parameter $m_\gs = 1/f$ which sets the scale 
of the $\gs$-model. The auxiliary fields $\hat{H}^\ga$ are 
defined as $\hat{H}^\ga = H^\ga - \gG^\ga_{\gb\gg} 
\bar{\psi}^\gb_R \psi^\gg_L$ with the connection 
$\gG^\ga_{\gb\gg} = G^{\ga\underline{\ga}} 
G_{\underline{\ga}\gb,\gg}.$ 

It is of particular importance to have an explicit 
expression for the kinetic terms of the Goldstone fields, 
which are modified by the presence of the matter terms 
in the K\"{a}hler potential (\ref{kp}). Following the 
procedures of \ct{jw1} and \ct{GNvH2}, the kinetic terms 
for the scalars $z^{ij}$ and the quasi-Goldstone 
fermions $\psi^{ij}_L$ are determined by the matter-extended 
K\"{a}hler metric 
\equ{
G_\gs(x, \bx) = {G_\gs}_{(ij)}{}^{(kl)} x^{ij} \bx_{kl} 
 = f^2 E \tr( x \gch^T \bx \gch ) + e^{f^2\, K_\gs} 
 f^2 k x \gch^T \bx \bk,
}
where $x^{ij}$ stands for components of the Goldstone 
superfield $\gF^{ij}$ or their gradients, whilst
\(
E =  \frac 1{2f^2} + e^{f^2\, K_\gs} k \gch\inv \bk 
- 2 e^{-2f^2\, K_\gs} | h |^2. 
\)
For some applications it is convenient to write this as
\equ{
G_\gs(x, \bx) = 
\tr( x \gch^T \bx \hat \gch), 
\qquad
\hat \gch = f^2 E \gch + e^{f^2 K_\gs} f^2 \bk k.
\label{chihat}
}
Clearly, the physical requirement that the model be 
ghost-free implies that this metric has to be sign-definite. 
As $\chi$ is positive definite, and the second term 
proportional to $\bar{k} k$ is non-negative, positive 
definiteness of the metric is guaranteed if $E > 0$. For 
$E < 0$, there always are negative kinetic-energy ghosts. 
However, for $E = 0$ a more detailed analysis is required. 

In particular, we note that 
\equ{
\det \hat \gch = f^{10} E^4 \left( E + e^{f^2 K_\gs} 
 k \gch\inv \bk \right) \det \gch. 
} 
As a result, for $E = 0$ the metric has a four-fold
zero eigenvalue. This implies the existence of four 
complex orthogonal eigenvectors of $\hat \gch$ with 
zero eigenvalue; one can then construct six independent 
(complex) anti-symmetric tensor zero-modes, of the form 
$x = v w^T - w v^T$, with $v, w$ independent zero 
eigenvectors. Of course, if $k = 0$ at the same time,
the whole metric $\hat{\chi}$ vanishes; then also 
the kinetic terms of all Goldstone fields and their
fermion partners vanish. 

As concerns mass-terms, we observe that for the 
model (\ref{kp}) it is not possible to construct 
an $SO(10)$-invariant superpotential. First, the 
non-linear transformations of the coordinates $z$ 
exclude their appearance in an invariant expression. 
Next, there is no non-vanishing holomorphic $SU(5)$ 
invariant for $k_i$. Finally, as $h$ transforms under 
$U(1)$ and there is no field that compensates for its 
transformation, it also cannot appear in the superpotential. 
In the absence of a superpotential, all fields in the 
action (\ref{so10l}) ---the Goldstone bosons and their 
superpartners as well as the chiral superfields defining 
the matter representations--- describe massless spin-0 
and chiral spin-1/2 particles. 

This situation changes if we add a second family of 
quarks and leptons, with superfields $\gS_{(2)} = 
(\gF_{(2)}^{ij}, \Psi_{(2)\, i}, \Psi_{(2)})$. It is 
then possible to construct an invariant superpotential 
\begin{equation} 
W(\gS)\, =\, \sum_{a = 1,2}\, \gl_a\, \Psi_{(a)}\, 
 \Psi_{(1)\, i} \Psi_{(2)\, j} \gF^{ij}_{(2)}. 
\label{W1}
\end{equation} 
The $\gl_a$ are coupling constants of dimension 
(mass)$^{-1}$. 

As a next step towards a physical interpretation of the 
fermions as describing quarks and leptons, we introduce 
gauge interactions. This can have important implications 
for the spectrum of the theory, as in supersymmetric 
theories gauge-couplings are accompanied by Yukawa 
couplings and a $D$-term potential. We first consider 
gauging the full $SO(10)$. A local transformation of the 
form (\ref{global}) then always allows one to go to the 
unitary gauge $z = \bar{z} = 0$. Thus all Goldstone 
bosons disappear from the spectrum as a result of the 
Brout-Englert-Higgs effect; this is confirmed by the 
finite mass-terms for the gauge fields corresponding 
to the broken generators of $SO(10)$.  

However in the presence of matter fields as in (\ref{kp}), 
required for the cancellation of anomalies, the analysis of 
the $D$-terms in the potential shows that in the unitary 
gauge the model becomes singular: in the minimum of the 
potential the expectation value of the K\"{a}hler metric 
vanishes: $E = 0$ and $k = 0$. Thus the kinetic energy 
terms of the Goldstone and quasi-Goldstone fields all 
vanish. Actually, this seems to happen in other fully 
gauged supersymmetric $\gs$-models on K\"{a}hler cosets 
with anomalies cancelled by matter as well. 

As an alternative to gauging $SO(10)$, one can gauge 
only the linear subgroup $SU(5) \times U(1)$ instead. 
This explicitly breaks the non-linear global $SO(10)$. 
It is then allowed in principle to construct 
superpotentials which are invariant only under the 
local gauge symmetry, although one would expect the 
strength of this potential to be proportional to the 
gauge coupling constant. In fact, this happens 
automatically with the $D$-term potentials. In addition, 
when gauging any group containing the $U(1)$ as a factor, 
the introduction of a Fayet-Iliopoulos term is allowed. 
It turns out, that the corresponding models are indeed 
well-behaved for a range of non-zero values of this 
parameter.  

We now present details of this analysis. The theory 
defined by the Lagrangian \eqref{so10l}, (\ref{kp}) has a 
global $SO(10)$ symmetry. This global symmetry allows vector 
bosons to be coupled to the model by turning the $SO(10)$ 
group, or its subgroup $SU(5)\times U(1)$, into a local 
gauge group by introducing covariant derivatives into the 
Lagrangian. The covariant derivatives are defined by:
\equ{
\mtrx{
\cD_\gm Z^\ga = \partial_\gm Z^\ga - A^i_\gm\cR^\ga_i, \\ 
 \cD_\gm\psi^\ga_L = \partial_\gm\psi^\ga_L - 
 A^i_\gm\cR^\ga_{i,\gb}\psi^\gb_L + 
 \cD_\gm Z^\gg\gG^\ga_{\gg\gb}\psi^\gb_L.
}
\labl{covariantD}
}
Here the $A_\gm^i$ are the gauge fields corresponding to 
the local symmetries. They are components of the vector 
multiplets $V^i = (A^i_\gm, \lambda^i, D^i)$, with 
$\lambda^i$ representing the gauginos and $D^i$ the real 
auxiliary fields. The isometries $\cR^\ga_i$ are generated 
by Killing vectors as in eq.\ (\ref{vKilling}); here they 
take the form  
\equ{
\gd_\gTh z = R, 
\qquad 
\gd_\gTh h = 2\tr (R_T) h, 
\qquad
\gd_\gTh k =  - k ( R_T  + \tr (R_T) \Id ),
\labl{vkilling}
}
with 
\(
R(z; \gTh)  = \frac 1f\,  x - u^Tz - zu + f\,  zx^\dag z 
\) and \(
R_T(z; \gTh) = -u^T + f\, zx^\dag.
\)
Adapting its normalization to that of the kinetic terms 
(\ref{kp}), the full Killing potential generating these 
Killing vectors is 
\(
\cM = \frac{1}{2} M_\gs + M_{(0;4)} + M_{(\bar{1};-4)}, 
\)
which takes the explicit form (cf.\ eq.(\ref{gD})):
\equ{
-i \cM = 
\tr \gD (-\frac 1{2f^2} - K_{(\overline{1};-4)} + 2  K_{(0;4)}) 
- e^{f^2 K_\gs} k\gD\gch^{-1}\bk.
} 
After introduction of the gauge fields in lagrangian 
\eqref{so10l}, via the covariant derivatives \eqref{covariantD}, 
the $\gs$-model itself is no longer invariant under 
supersymmetry transformations. Supersymmetry is restored by 
adding terms
\equ{
\gD\cL_\cK = 2G_{\ga\underline{\ga}}
(\cR^\ga_i\bar{\psi}^{\underline{\ga}}_L\lambda^i_R + 
\bar{\cR}_i^{\underline{\ga}}\bar{\lambda}^i_R\psi^\ga_L) - 
D^i(\cM_i + \gx_i).
\labl{LD}
}
We have added a Fayet-Iliopoulos term with parameter $\gx_i$ in 
case there is a commuting $U(1)$ vector multiplet. The full 
lagrangian for this model after introducing gauge interactions 
becomes 
\equ{
\cL=  \cL_{YM} + \cL_{chiral},
}
where $\cL_{YM}$ is the usual supersymmetric Yang-Mills action,
of the generic form 
\equ{
\cL_{YM} = -\, \frac{1}{g^2}\, \Tr (\frac{1}{4}\cF^2_{\gm\gn}  
 + \frac{1}{2} \bar{\lambda} D\Slashed \lambda - 
 \frac{1}{2}D^2 ).
\labl{LYM}
}
We use $\Tr$ to denote a trace over $2N\times 2N$-matrices; in 
contrast, traces over $N \times N$-matrices are denoted by tr.
When gauging a product of several commuting subgroups of $G$, 
e.g.\ $SU(5) \times U(1)$, there is a coupling constant $g_i$ for 
each of the subgroup factors. $\cL_{chiral}$ is given by 
\eqref{so10l}, but with ordinary derivatives $\partial_\gm Z^\ga$, 
$\partial_\gm\psi^\ga_L$ replaced by the covariant derivatives 
\eqref{covariantD}, while adding $\gD\cL_\cK$:  
\equ{ 
\cL_{chiral} = \cL_\cK(\partial_\gm\rightarrow \cD_\gm) + 
\gD\cL_\cK .
\labl{Lmatter}
}
Next we analyze the scalar potential obtained by elimination
of the $D$-fields for various gaugings. By substituting the 
expression \eqref{gD} for $\gD$ we obtain in index-free 
notation\footnote{The factors $-i$ here result from $\Delta$ 
being anti-hermitean, eq.(\ref{MgD}).} 
\equa{
-i\cM_u & = (\Id - 2f^2 \bz\gch z ) 
\Bigl(\frac1{2f^2}  + K_{(\overline{1};-4)} - 2  K_{(0;4)} 
 \Bigr) + e^{f^2 K_\gs}(k^T\bk^T - f^2 \bz\bk k z),
\non \\[2mm]
-i\cM_{x^\dag} & = 
-f \bz\gch \Bigl(-\frac 1{2f^2} - K_{(\overline{1};-4)} + 2  
 K_{(0;4)} \Bigr) + f e^{f^2 K_\gs}\bz\bk k,
\labl{so10killing} \\[2mm]
-i\cM_x & = f \gch z \Bigl(-\frac 1{2f^2}- K_{(\overline{1};-4)} 
 + 2  K_{(0;4)} \Bigr) - f e^{f^2 K_\gs}\bk k z.
\non 
\labl{Kilpot}
}
If the full $SO(10)$ is gauged, the unitary gauge can be 
chosen in which all Goldstone bosons $(z,\bar{z})$ vanish. 
This implies that the broken Killing potentials $\cM_x$ 
and $\cM_{x^\dag}$ vanish automatically, leaving us with 
the $U(5)$ Killing potentials only. If we only gauge 
$U(5)$ then the Killing potentials $\cM_x$ and $\cM_{x^\dag}$ 
are irrelevant, and again we have to consider only the $U(5)$ 
Killing potentials. However, in this case $z$ represents a 
physical degree of freedom, and its vacuum expectation 
value does not necessarily vanish: $\langle z \rangle = 0$ 
is guaranteed only if $SU(5)$ is not broken. 

To analyze both gauged $SO(10)$ and gauged $SU(5) \times U(1)$ 
at once, we consider the $D$-term potential arising from the 
gauging of $SU(5)\times U(1)$ including a Fayet-Iliopoulos 
term with parameter $\gx$ for the $U(1)$:
\equ{
V = \frac {g_1^2}{2 N} (\gx -i \cM_Y)^2 
+ \frac{g_5^2}{2} \tr (-i \cM_t)^2.
\labl{ScalarPot}
}
Here the $U(5)$ Killing potentials $\cM_Y$ and $\cM_t$ are 
trace and the traceless part of $\cM_u$: 
\equ{
\cM_t = \cM_u - \frac 1N \cM_Y \, \Id, 
 \qquad \cM_Y = \tr \cM_u.
}
We can derive $\tr \cM_t^2$ from $\cM_Y$ and $\tr \cM_u^2$ by 
\equ{
\tr (-i\cM_t)^2 = \tr (-i \cM_u)^2 - \frac 1N (-i\cM_Y)^2.
}
An explicit expression for $\cM_t$ in terms of the 
matter-extended submetric $\hat{\chi}$ is 
\equ{
- i \cM_t = \frac{2}{f^2} \hat \gch^T - 2 \gg \Id 
 - e^{f^2 K_\gs} \left( f^2 \bz \bk \, k z + k^T \bk^T
\right),
}
where $\gg$ is defined by 
\begin{equation} 
N \gg = (\tr \gch) E + \frac 1{2} 
 e^{f^2 K_\gs} k( 1 - f^2 z \bz ) \bk.
\end{equation}
The terms in the potential (\ref{ScalarPot}) are proportional 
to the square of the coupling constants $g_1$ and $g_5$ of 
the $U(1)$ and $SU(5)$ gauge groups, respectively. The case 
of fully gauged $SO(10)$ is reobtained by taking the coupling 
constants equal: $g_1 = g_5 = g_{10}$, and the 
Fayet-Iliopoulos term to vanish: $\gx = 0$. We have left the 
rank $N = 5$ of $SO(10)$ in, so as to keep track of some of 
the dependence on this rank.  

The potential (\ref{ScalarPot}) is non-negative. In 
order for supersymmetry to be preserved, the minimum must 
be at $V_{min} = 0$; in contrast, $V_{min} > 0$ implies 
spontaneous supersymmetry breaking by the potential. 
Being a sum of squares, a vanishing potential is possible 
only if $\cM_Y = 0$ and $\cM_t = 0$ at the same time. 

In the case of gauged $SO(10)$ one can always work in 
the unitary gauge $z = \bar{z} = 0$. However, in the
case of gauged $SU(5) \times U(1)$ the potential can 
cause further symmetry breaking by generating a 
vacuum expectation value for the would-be Goldstone 
bosons. Because of its antisymmetry, an $SU(5) \times 
U(1)$ transformation can be performed to put $\langle 
z \rangle$ into the standard form 
\equ{
\langle f z \rangle = \pmtrx{
 a \gs_2 & & \\
         & b \gs_2 &  \\
& & 0
}, 
\label{zvev} 
}
with real $a,b \geq 0$. Of course, the unitary gauge is
included as the special case $a = b =0$. The vacuum 
expectation value (\ref{zvev}) preserves a subgroup $SU(2) 
\times SU(2) \times U(1)$. If the $\bar{\underline{5}}$ 
gets a vacuum expectations value, this residual symmetry 
can be used to chose 
\begin{equation} 
\langle k \rangle = \lh k_1, 0, k_3, 0, k_5 \rh. 
\label{kvev} 
\end{equation} 
We first investigate the existence of zeros of the potential,
compatible with supersymmetry. The condition $\langle \cM_t 
\rangle = 0$ then implies $k_1 = k_3 = 0$, and 
\begin{equation} 
E = \gg \lh 1 + a^2 \rh = \gg \lh 1 + b^2 \rh = 
 \gg - \lh 1 + a^2 \rh \lh 1 + b^2 \rh |k_5|^2. 
\label{susE}
\end{equation} 
There are three separate solutions to these conditions; the 
first is 
\begin{equation} 
a = b = k_5 = 0, \hspace{2em} E = \gg.
\label{unisol}
\end{equation} 
This solution includes the unitary gauge. A second solution 
(which coincides with the previous one for $a = b =0$) is the 
case $E = \gg = k_5 = 0$. It can be seen immediately to yield 
$\hat{\chi} = 0$. Therefore in this case the kinetic terms of 
the Goldstone superfield components vanish. Such a solution 
is unacceptable, not only because part of the quarks and leptons 
disappear from the spectrum of physical states, but even 
more importantly as this upsets the cancellation of anomalies,
which is guaranteed only if all chiral fermions in the model 
contribute. This holds in particular for the case of
fully gauged $SO(10)$ in the unitary gauge. 

The third solution of the supersymmetric vacuum conditions,
which exists only for $E, \gg < 0$, is 
\begin{equation} 
a^2 = b^2 = - \frac{1}{\gg} \lh 1 + a^2 \rh^4 |k_5|^2, 
 \hspace{2em} E = \gg \lh 1 + a^2 \rh. 
\label{nonunisol}
\end{equation} 
Inserting this solution into the expression (\ref{chihat})
one obtains $\hat{\chi} = f^2 \gg \Id$, which in this case 
is negative definite. As it is not possible to change 
the overall sign of the K\"{a}hler potential without 
creating negative kinetic energy terms for the matter 
fields, this solution always contains ghosts and is again 
physically unacceptable. 

The upshot of this discussion is, that physically consistent 
models (i.e.\ anomaly-free, with positive definite kinetic 
energy), in which the potential has zeros, require  
$z = k = 0$ and $E = \gg > 0$. Such models can be realized 
with gauged $SU(5) \times U(1)$, but the model with fully 
gauged $SO(10)$ is excluded. We observe, that positivity 
of $E$ for these solutions implies 
\begin{equation}  
0 \leq |h|^2 < \frac{1}{4f^2}.
\label{hvev}
\end{equation} 
Thus, unless $h = 0$, these solutions always spontaneously
break $U(1)$, whilst $SU(5)$ is manifestly preserved. 
 
The conditions (\ref{unisol}) and (\ref{hvev}) are necessary
to have physically consistent models with $\langle \cM_t 
\rangle = 0$. This is sufficient for a zero of the potential 
in a model with gauged $SU(5)$ only. If $U(1)$ is gauged, 
a zero of the potential requires the additional condition: 
\begin{equation} 
\begin{array}{ll}
\xi = \langle i\cM_Y \rangle\, = & \dsp{ 
 - \lh 1 + a^2 \rh^2 \lh 1 + b^2 \rh^2 \left[ 
 \lh 1 - a^2 \rh |k_1|^2 + \lh 1 - b^2 \rh |k_3|^2 + 
 |k_5|^2 \right] }\\ 
 & \\ 
 & \dsp{ - \lh 1 + 2 \frac{1 - a^2}{1 + a^2} + 
 2 \frac{1 - b^2}{1 + b^2} \rh\, E.}
\end{array} 
\label{zeroY}
\end{equation} 
Combining this with $z = k = 0$, it follows that (with 
$N = 5$) 
\begin{equation} 
E  = \gg = - \frac{\xi}{N}, \hspace{2em} 
 |h|^2 = \frac{1}{4f^2} + \frac{\xi}{2N}. 
\label{susEh}
\end{equation} 
A consistent solution of this type exists only for 
$- N/(2f^2) \leq \xi < 0$. The kinetic energy for the 
Goldstone superfield components is now proportional to 
$(- f^2 \xi)/N$.

Clearly, for values of $\xi$ in this range it is necessary 
to perform a finite renormalization of the Goldstone 
superfields to obtain the canonical value of the kinetic 
terms; in the K\"{a}hler potential this is equivalent to 
a rescaling of the $\gs$-model scale such that $f^2 
\rightarrow -N/\xi$. In these models the natural value of 
the Fayet-Iliopoulos-parameter is therefore the $\gs$-model 
scale, thereby relating internal and supersymmetry breaking. 

We finish this section by observing that, in addition to 
zeros of the potential, there can also be ranges of the 
parameters $(g_1^2, g_5^2, \xi)$, or models with only 
some proper subgroup of $SU(5)$ gauged, for which the 
minimum of the potential occurs at a positive value: 
$\langle V \rangle > 0$. In this case supersymmetry is 
manifestly broken by the potential. This could happen 
for example in the domain $\xi < - N/(2f^2)$. However, 
we have not performed an exhaustive analysis of this case.

%
%
\section{Conclusion}
\label{Conclusion}

\nit
In this paper we have considered supersymmetric models based 
on classic \Kh ian coset spaces: $U(M+N)/U(M)\times U(N), 
USp(2N)/U(N)$ and $SO(2N)/U(N)$, and their non-compact 
versions. Starting from a non-linear realization of the group 
$SL(N+M, \Cplx)$ in finite form, we constructed their \Kh\ 
potentials. A generalization of the Killing potential for 
finite transformations has been obtained. The \Kh\ potential 
of such a coset can be written as a function of a fundamental 
submetric. This submetric also allows us to construct \Kh\ 
potentials for superfields as  sections of bundles over the 
original classical coset. For most of these matter 
representations the naive definitions are sufficient to 
guarantee the existence of these bundles globally. However, 
the consistency of line bundles requires that the cocycle 
condition is satisfied. 

We have discussed various aspects of these general 
constructions for classical \Kh ian coset space in more 
detail for the class of orthogonal cosets $SO(2N)/U(N)$. 
All supersymmetric matter fields which form completely 
anti-symmetric representations of $SU(N)$ with arbitrary 
integer charges satisfying the cocycle condition have 
been obtained explicitly. 

Pure supersymmetric coset models are often anomalous due to 
their chiral fermions. This is also the case for orthogonal 
cosets $SO(2N)/U(N)$, but  as all $SO(2N)$ representations 
are anomaly free (with the exceptions of $SO(2) \cong U(1)$ 
and $SO(6) \cong SU(4)$), the supersymmetric field content 
can be extended such that all anomalies cancel. 
The completely anti-symmetric $SU(N)$ representations 
descending from the positive-chirality spinor representation 
of $SO(2N)$ provide possible candidates for anomaly free 
models, which can include the Goldstone bosons. 
However, the $U(1)$ charges of these anti-symmetric 
representations can often not be realized using the bundles 
at our disposal. In fact, only for $N = 2, 5, 6, 8$ these 
$SO(2N)/U(N)$-spinor models can fulfil the consistency 
requirements of the line bundle. 

Some phenomenological aspects of the $SO(10)/U(5)$-spinor 
model have been investigated. This model contains the 
$SU(5)\times U(1)$ fermionic field content of one generation 
of quarks and leptons, including a right-handed neutrino. 
The matter-extended metric for the Goldstone bosons of the 
coset is not automatically positive definite. In order that 
the theory is ghost-free when expanded around a minimum of 
the potential, the quantity $E$ has to be positive, see 
eq.\ \eqref{chihat}. The consequences of this physical 
requirement have been analysed for supersymmetric minima, 
if part of the isometry group is gauged. If the whole 
$SO(10)$ is gauged, the analysis is straightforward as one 
can employ the unitary gauge to put the Goldstone bosons to 
zero. We find the kinetic energy of the would-be Goldstone 
modes and their fermionic partners to vanish. Therefore the 
quasi-Goldstone fermions no longer contribute to the 
cancellation of anomalies. 
 
Gauging (part of) the linear subgroup $U(5)$ calls for 
a more involved investigation. First we have obtained 
all supersymmetric minima for the case where $SU(5)$ is 
gauged. We found three classes of such vacua, of which two 
are physically problematic as the kinetic terms of the 
Goldstone multiplets either vanish or have negative values. 
The third type of supersymmetric vacuum only exists for a 
finite range of vacuum expectation values of the scalar 
partner of the right-handed neutrino. If the $U(1)$ factor 
is gauged in addition, the Fayet-Iliopoulos parameter is 
related directly to the vacuum expectation value of this 
scalar. This shows that only for a finite range of values 
of the Fayet-Iliopoulos parameter $U(5)$ can be gauged 
consistently. 

%
%
\appendix 
%
%
\section{Decomposition of $\boldsymbol{SO(2N)}$ spinors
into anti-symmetric tensors}
\label{DecompSpin}

An arbitrary spinor $\Bgps$ of $SO(2N)$ can be represented 
using anti-symmetric tensors  $\gps_{p\, i_1 \ldots i_{p}}$ 
of $SU(N)$ with $p$ indices as 
\equ{
\Bgps = \lh \gps_0, \gps_{1}^{\,  i_1}, \ldots , 
\gps_{N}^{\, i_1 \ldots i_{N}} \rh.
}
The invariant inner-product of two spinors $\Bgps$ and $\Bgf$ 
is given by
\equ{
\Bgps^\dag \Bgf  = 
\sum_{p=0}^N \frac 1{p!} 
\gps^\dag_{p\, i_{p} \ldots i_{1}}
\gf_{p}^{~  i_1 \ldots i_{p}}
\label{SpinInner}
}
where 
\(
\gps^\dag_{p\, i_{p} \ldots i_{1}} = 
\gps_{p}^{*\, i_{1} \ldots i_{p}}
\). 
We want to construct a basis for the anti-symmetric 
$SU(N)$-tensors, and also a basis for the $SO(2N)$-spinors, 
using the Clifford algebra of fermion creation and annihilation 
operators $\gG^i$ and $\bgG_i$, as introduced by R.N.\ Mohapatra 
and B.\ Sakita \cite{MS}, see also \cite{Polchinski} and 
\cite{GSW-II}. They satisfy the usual anti-commutation 
relations 
\equ{
\{\gG^i,\bgG_j\} = \gd^i\,_j ,
\qquad 
\{\gG^i,\gG^j\} = \{\bgG_i,\bgG_j\} = 0
\labl{Gr}.
}
Assume that we have constructed a Hilbert space on which these 
Clifford operators act. In this Hilbert space we define the 
vacuum state $| 0 \rangle$ by $\gG^i | 0 \rangle = 0$ for any 
$i$. The ket- and bra-states 
\equ{
{\bf e}_{p\, i_1 \ldots i_p}   = 
\bgG_{i_1} \ldots \bgG_{i_p} | 0 \rangle
\qquad 
{\bf e}_p^{\dag\, i_p \ldots i_1}  = 
\langle 0 | \gG^{i_p} \ldots \gG^{i_1}
}
satisfy the orthonomality relations 
\equ{
{\bf e}_p^{\dag\, i_p \ldots i_1} {\bf e}_{q\, j_1 \ldots j_q}  
 = 0, \quad \text{for} \quad p \neq q 
\qand 
{\bf e}_p^{\dag\, i_p \ldots i_1} {\bf e}_{p\, j_1 \ldots j_p} 
 = \gd^{i_1}_{[j_1} \ldots \gd^{i_p}_{j_p]}, 
}
where $\gd^{i_1}_{[j_1} \ldots \gd^{i_p}_{j_p]}$ is the 
complete anti-symmetrized Kronecker-delta. Therefore the 
states ${\bf e}_{p\, i_1 \ldots i_p }$ form a basis of 
anti-symmetric rank $p$ tensors of $SU(N)$. Using the 
complete anti-symmetry it is easy to show that the number 
of the vectors ${\bf e}_p$ with length $p$ is equal to 
$\smash{\binom N p}$, hence the total number of vectors 
$\{ {\bf e}_p\}$ is equal to $2^N$. The collection of these 
states ${\bf e}_p$ for $0 \leq p \leq N$ form a basis for 
$SO(2N)$-spinors, hence $\Bgps$ and $\Bgps^\dag$ can be 
expanded in this basis 
\equ{
\Bgps =  
\sum_{p=0}^N \frac 1{p!} 
\gps_{p}^{~  i_{1} \ldots i_{p}}
{\bf e}_{p\, i_1 \ldots i_p} 
\qand
\Bgps^\dag =  
\sum_{p=0}^N \frac 1{p!} 
\gps^\dag_{p\, i_{p} \ldots i_{1}}
{\bf e}_{p}^{\dag\,  i_1 \ldots i_p}.
}
It is straightforward to check that in this basis the 
inner-product of two spinors $\Bgps^\dag \Bgf$ is consistent 
with the definition \eqref{SpinInner} using the Clifford 
properties \eqref{Gr}.

In terms of the Clifford algebra, we define the $2N$ 
gamma-matrices $\gG_a$ with $a = 1,\dots, 2N$ by 
\equ{
\gG_a =
\begin{cases}
i(\gG^i - \bgG_i),& a = i = 1,\dots, N, 
\\[2mm]
~~ \gG^i + \bgG_i,& a = i +N = N+1,\dots, 2N,
\end{cases}
}
with the property 
\equ{
\{\gG_a,\gG_b\} = 2\gd_{ab}.
}
This property can be used to show that the sigma-matrices 
\equ{
M_{ab} = \frac{1}{2}\gS_{ab}  =\frac{1}{4}[\gG_a,\gG_b],
}
are the generators of the $SO(2N)$-algebra \eqref{so} 
in the spinor representation. With respect to the spinor 
inner-product \eqref{SpinInner} the gamma-matrices are 
Hermitean $\gG^\dagger _a = \gG_a$ and hence the 
sigma-matrices are anti-Hermitean $\gS^\dagger _{ab} = 
- \gS_{ab}$. Furthermore it implies that w.r.t.\ this inner 
product the fermion creation/annihilation operators are 
Hermitean conjugates: 
\equ{
\lh \gG^i \rh^\dag = \bgG_i, 
\qquad
\lh \bgG_i \rh^\dag = \gG^i.
}
For products of Clifford operatros $A$ and $B$ we have 
$(AB)^\dag = B^\dag A^\dag$. 

The Hermitean chirality operator $\tgG$ defined by
\equ{
\tgG = (-)^{\half N(N-1)}i^{-N} \prod_{a=1}^{2N} \gG_a 
}
can be written in terms of the Clifford elements as
\equ{
\tgG = \prod_i [ \gG^i, \bgG_i ] = 
\prod_{i = 1}^N (1 - 2 \hat n_i) = (-)^{\hat n}.
\label{chirality}
}
Here we have defined the $i$th number operator $\hat n_i = 
\bgG_i \gG^i$ and the total number operator $\hat n = \sum_i 
\hat n_i$. Using this chirality operator, we can define 
positive and negative chirality spinors in $2N$ dimensions
\equ{
\tgG \Bgps_\pm = \pm \Bgps_\pm
}
Using the form of the chirality operator \eqref{chirality}, 
it follows that the positive chirality components of a spinor 
are given by completely anti-symmetric $SU(N)$-tensors of even 
length $p$, while the negative chirality components have odd 
length $p$. 

\nit
The generators $U^i\,_j$ can be expressed in terms of the 
fermion operators as 
\equ{
U^i\,_j = - \frac{1}{2}[\gG^i,\bgG_j],
} 
and satisfy the $U(N)$ algebra \eqref{un}. Their anti-symmetric 
part $A^i\,_j$, and symmetric part $S^i\,_j$, take the form
\equ{
\mtrx{
A^i\,_j = - \frac{1}{4}\left([\gch^i,\bar{\gch_j}]- 
[\gch^j,\bar{\gch_i}]\right)
,&S^i\,_j = \frac{i}{4}\left([\gch^i,\bar{\gch_j}]+ 
[\gch^j,\bar{\gch_i}]
\right)
}
\non
}
Furthermore the broken $SO(2N)$ generators $X^{ij}$ and 
$\bX_{ij}$ can be represented by
\equ{
X^{ij} = \gG^i \gG^j
\qand
\bX_{ij} = \bgG_i \bgG_j.
\non
}
The $U(1)$-charge operator \eqref{u1} is given in terms of 
the total number operator $\hat n$ by 
\equ{
Y = \sum_i [ \gG^i, \bgG_i ] = N - 2 \hat n,
\label{Ycharge}
}
hence the charge of an anti-symmetric tensor with $p$ 
indices, that occurs in the decomposition of a spinor, is 
$N - 2p$:
\(
Y {\bf e}_p = (N - 2p) {\bf e}_p.
\)
We define the dual vectors ${\bf e}_{\sscp \ovr{N-p}}$ resp. 
${\bf e}^\dag_{\sscp \ovr{N-p}}$
of the basis vectors ${\bf e}_p$ and ${\bf e}^\dag_p$ resp.\ 
by
\equ{
{\bf e}_{\sscp \ovr{N-p}}^{~~~~ i_N\ldots i_{p+1}} = 
\frac 1{p!} \ge^{i_N \ldots i_1} 
{\bf e}_{p\, i_1 \ldots i_p}
\qand 
{\bf e}^\dag_{\sscp \ovr{N-p}\,  i_{p+1}\ldots i_{N}} = 
\frac 1{p!}  
{\bf e}_{p}^{\dag\, i_p \ldots i_1} \ge_{i_1 \ldots i_N}.
\labl{DualOp}
}
For the components $\gps_p$ we use analogous definitions.
Notice that under dualization the charge does not change, 
only the number of indices does. 

%
%
\section{Anomaly cancellation of the spinor representation}
\label{AnomalyCancellation}

We now show that the positive chirality spinors of $SO(2N)$ 
have no pure $U(1)$-anomaly, unless $SO(2N)$ is isomorphic 
to a non-anomaly-free unitary group, $SO(2) \cong U(1)$ 
or $SO(6) \cong SU(4)$, by computing the possible 
$U(1)$-anomalies. However it is straightforward to also 
compute the $U(1)$-anomalies for negative chiralities, so we 
calculate both here. The $Y^k$-anomaly $A_\pm(Y^k;N) = 
\Tr_\pm Y^k$ for the $\pm$ chirality spinor representation 
is given by
\equ{
A_\pm (Y^k;N) = 
\sum_{l=0}^N \binom Nl \frac {1 \pm (-)^l}2 (N - 2l)^k.
\labl{anomalyYk}
}
This follows using the multiplicities $\smash{\binom Nl}$ 
and charges $N-2l$ of the states $\gps_l^{~i_1\ldots i_l}$. 
The factor $\smash{ \frac {1 \pm (-)^l}2 }$ is introduced 
to project onto the positive or negative chirality states. 
The necessary details to obtain these results can be found 
in appendix \ref{DecompSpin}. To calculate these anomalies 
it is convenient to introduce the functions 
\equ{
q_\pm (x) = \pm \frac 1x + x
}
of a variable $x$, in terms of which we define
\equ{
P_\pm(x;N) \equiv \half \left[ (q_+)^N \pm (q_-)^N \right] 
 = \sum_{l=0}^N \binom Nl \frac {1 \pm (-)^l}2 x^{N-2l}.
} 
Notice that the charge operator $Y$ can be represented by 
\(
Y = x \frac d{dx}
\).
The anomaly $A_\pm(Y^k)$ can be calculated using the functions 
$P_\pm$ by
\equ{
A_\pm(Y^k;N) = 
\left.
\lh x \frac d{dx} \rh^k P_\pm(x;N)
\right|_{x = 1}.
}
To compute this we use the properties of the functions $q_\pm$ 
\equ{
x \frac d{dx} q_\pm = q_\mp, \quad
q_+|^{}_{x = 1} = 2, \quad \text{and} \quad 
q_-|^{}_{x = 1} = 0. 
}
We obtain the following results for the $Y$ and $Y^3$ 
anomalies in $D = 4$ dimensions
\equ{
A_\pm (Y; N) = 
\begin{cases}
\pm 1 & N = 1, \\
0     & N \neq 1,
\end{cases}
}
and 
\equ{
A_\pm (Y^3; N) = 
\begin{cases}
\pm 3! 2^2 & N = 3, \\
\pm 1      & N = 1, \\
0          & N \neq 1,3.
\end{cases}
}
Hence we see that the cases $N=1$ and $N = 3$ have indeed 
an anomalous spinor representation. We conclude from this 
anomaly analysis that for $N = 2$ and $N \geq 4$ the spinor 
representation of $SO(2N)$ is $U(1)$ anomaly-free.

\end{document}